\documentclass[11pt,draftcls,onecolumn]{IEEEtran}
\usepackage{epsfig, multirow, makecell, array, cite}
\usepackage{amssymb, amsmath}
\usepackage{epstopdf,subfigure}
\usepackage{color}
\ifCLASSINFOpdf
\else
\fi
\def\a{a}
\def\b{b}
\def\c{c}

\begin{document}
\newcommand\comb[2]{\ensuremath\vphantom{\mathrm C}_{#1}
        \kern -.1em\mathrm{C}\kern -.09em{}_{#2}}

\title{The Degrees of Freedom of the Interference Channel with a Cognitive Relay under Delayed Feedback}
\author{\IEEEauthorblockN{Hyo Seung Kang, \textit{Student Member, IEEE}, Myung Gil Kang, \textit{Student Member, IEEE}, Wan Choi, \textit{Senior Member,
IEEE}, and Aria Nosratinia,~\IEEEmembership{Fellow, IEEE}}
\thanks{H. S. Kang, M. G. Kang and W.~Choi are with Department of Electrical
    Engineering, Korea Advanced Institute of Science and Technology
    (KAIST), Daejeon 305-701, Korea (e-mail: \{khs5667, casutar\}@kaist.ac.kr,
    wchoi@kaist.edu). A. Nosratinia is with the Department of Electrical Engineering at the University of Texas at Dallas, TX, 75023
    U.S.A., e-mail: aria@utdallas.edu}
}
\maketitle
\newtheorem{lemma}{Lemma}
\newtheorem{theorem}{Theorem}
\newtheorem{definition}{Definition}
\newtheorem{corollary}{Corollary}
\newtheorem{remark}{Remark}
\newtheorem{example}{Example}
\allowdisplaybreaks
\vspace{-0.7in}
\begin{abstract}
This paper studies the
interference channel with a cognitive relay (ICCR) under delayed feedback.
Three types of delayed feedback are studied: delayed channel state
information at the transmitter (CSIT), delayed output feedback,
and delayed Shannon feedback. Outer bounds are derived for the DoF region
of the two-user multiple-input multiple-output (MIMO) ICCR with
delayed feedback as well as without feedback. For the single-input
single-output (SISO) scenario, optimal schemes are proposed based on
retrospective interference alignment. It is
shown that while a cognitive relay {\em without} feedback cannot extend the sum-DoF beyond $1$  in the two-user SISO interference
channel, delayed feedback in the
same scenario can extend the sum-DoF to $4/3$. For the MIMO case,
achievable schemes are obtained via extensions of retrospective
interference alignment, leading to DoF regions that meet the
respective upper bounds.
\end{abstract}

\section{Introduction}\label{intro}
Cognitive radio is a subject of intense interest motivated by its
potential for better usage of spectral resources. To explore the
fundamental limits of such channels, and to make use of powerful
techniques developed for capacity of channels with state known at
transmitter, some information-theoretic cognitive models allow a
cognitive node to possess non-causal knowledge of data originating
elsewhere. Interestingly, in the recent years  applications
have emerged where knowledge of another nodes' data prior to
transmission is indeed practically viable. Examples include
heterogeneous networks or coordinated networks, where some base
stations can possess knowledge of the messages of other base stations
by coordination. Other examples involve layered cell structures, where
macro base stations can know the messages of pico base stations that
are routed from the macro base station over backhaul links. Such
heterogeneous or coordinated networks can be modeled by interference
channels with cognitive transmitters~\cite{KC2012}.

Contrary to the model in \cite{KC2012}, when a cognitive transmitter does not have its intended receiver, it is called cognitive relay and helps other transmitters in a way of reducing the effective interference at the receivers. In this paper, we consider the interference channel
with a cognitive relay (ICCR)\footnote{It is also known by the name cognitive relay-assisted interference channel. } where transmitter-receiver pairs constitute an interference channel and the cognitive relay helps the transmitters.

Previous works in this area have generally focused on perfect and instantaneous channel state
information at transmitter (CSIT). However, feedback delays are often present in real systems and make feedback information outdated. Fortunately, the usefulness of delayed CSIT has been explored in various channel models \cite{MAT2012, VV2011, AGK2011,MJS2012,AGK2011_2,GMK2011,KC2013,VV2012,TMPS2013,AGK2011_3,MVV2012} . The ICCR with delayed feedback, nevertheless, has not received much attention despite its importance.



\subsection{Past Work}\label{pase_work}
The ICCR was first considered in~\cite{SE2007} where an achievable rate region
via a combination of dirty paper coding \cite{C1983} and beamforming was reported.
In \cite{SVJS2008}, a new achievable region was
presented by a combination of the Han-Kobayashi coding scheme \cite{HK1981}
and dirty paper coding, and an outer bound for the Gaussian ICCR was derived.
For a discrete memoryless (DM) ICCR, an outer bound was first derived
in \cite{RTD2010} and then improved achievable rate regions and outer bounds
were reported in~\cite{RTDG2011, RTDG2011_2, CHN2013,DRDT2014}.
The capacity region of DM-ICCR is known in very strong and
strong interference regime \cite{RTDG2011, RTDG2011_2}, but it still remains
unknown under general channel conditions.

When capacity remains intractable, the degrees of freedom (DoF) are
often used to understand the asymptotic characteristics of the
capacity. The DoF is defined as the ratio of the capacity of the
channel of interest to a simple SISO Gaussian channel capacity, when
transmit power goes infinity. The DoF of ICCR has been studied
in~\cite{SVJS2008, CS2012, WS2012}. It was proved in~\cite{SVJS2008}
that the two-user Gaussian ICCR has DoF $2$ almost surely if perfect
and instantaneous CSIT and
CSI at receiver (CSIR) are available, which implies that each receiver
does not suffer from interference in an asymptotic sense. For the
$K$ users with perfect CSIT and CSIR, achievable sum DoF and outer
bounds of the sum DoF were derived in~\cite{CS2012, WS2012}. Although
a conventional relay cannot increase the DoF~\cite{TN2012}, the authors
of~\cite{WS2012} showed that a cognitive relay can improve DoF unlike
a conventional relay; the optimal sum DoF for $K$ users with perfect
CSIT is $\frac{K+1}{2}$ if $K$ is odd while the sum DoF for the
$K$-user interference channel with perfect CSIT is $\frac{K}{2}$ by
interference alignment~\cite{CJ2008}.

The usefulness of delayed CSIT has been first demonstrated in \cite{MAT2012}
for multiple-input single-output broadcast channel (BC). In \cite{MAT2012},
the base station exploits the delayed CSI to estimate the interference at each receiver in the previous transmission (i.e., the side information at the receivers) and then retrospectively align the interfering signals with the help of the side information. For multiple-input
multiple-output (MIMO) BC with delayed CSIT, an outer
bound of the DoF region with $K$ users and the DoF region with two
users were derived in~\cite{VV2011}, and sum DoF for a three-user case
was obtained in~\cite{AGK2011}. New retrospective interference
alignment schemes for an interference channel and an $X$ channel with
delayed CSIT and delayed output feedback were proposed
in~\cite{MJS2012}, and the sum DoF was derived.  The achievable DoF
reported in~\cite{MJS2012} was improved
in~\cite{AGK2011_2,GMK2011}. Recently,~\cite{KC2013} proved the
usefulness of ergodic interference alignment in a $K$-user
interference channel when only delayed feedback is available and
showed that the sum DoF of 2 can be achievable as $K$ goes to
infinity, which is the best DoF result until now in a $K$-user
interference channel with delayed CSIT.
derived in~\cite{VV2012, TMPS2013,AGK2011_3}.
In~\cite{VV2012}, the DoF region with delayed CSIT was derived
for general MIMO interference channel  with an arbitrary numbers of antennas.
The authors of \cite{VV2012} showed that Shannon feedback, which has both output feedback and delayed CSIT,
strictly enlarges the DoF region of the MIMO interference channel compared to the case with delayed CSIT only \cite{TMPS2013}. For delayed local CSIT, an achievable DoF region of MIMO interference channel was derived in~\cite{AGK2011_3}.
In~\cite{MVV2012}, the authors presented a
hybrid CSIT model where one transmitter has perfect and instantaneous
knowledge of channel matrices corresponding to one user while the
other transmitter has only delayed CSI corresponding to the other
user, and derived the DoF region of the MIMO interference channel with
hybrid CSIT.  Moreover, the DoF regions of MIMO interference channel
and broadcast channel without CSIT were derived in~\cite{HJSV2012},
and in addition to MIMO interference channel and broadcast channel,
the DoF region of a cognitive radio channel without CSIT was reported
in~\cite{VV2012_2}.

\subsection{Main contribution}\label{contribution}
In this paper, we consider the interference channel with cognitive relay in the presence of various types of delayed
feedback at the transmitter in independent
and identically distributed (i.i.d.) fading
channels. In all cases perfect CSIR is assumed. Unless explicitly
mentioned otherwise, a two-user system is considered. The types of
delayed feedback information (including no feedback) are
\begin{itemize}
\item Delayed CSIT: Transmitters and a cognitive relay know all channels after one sample delay.
\item Delayed output feedback: Transmitters and a cognitive relay know output of their intended receiver after one sample delay.
\item Delayed Shannon feedback: Transmitters and a cognitive relay know all channel gains and the output of intended receiver after one sample delay.
\item No feedback: Both transmitters and a cognitive relay do not have any channel information.
\end{itemize}

For each type of delayed feedback, an outer bound for the DoF region is derived.
Focusing on the special case of the single-input
single-output (SISO), a scheme is proposed that achieves the outer bound based on the
retrospective interference alignment for each type of feedback. From
the derived DoF region, it is shown that the sum DoF in the
single-antenna network is $\frac{4}{3}$ for delayed CSIT, delayed
output feedback, and delayed Shannon feedback with the help of a
cognitive relay, compared with the sum-DoF of the interference channel
which is only 1 regardless of availability of CSIT. It is also shown
that a cognitive relay does not extend the DoF region in the absence
of CSIT.

The proposed retrospective interference alignment scheme is
extended to the MIMO case.
It is shown that for the three types of delayed feedback
information, the DoF regions achieved by the proposed retrospective
interference alignment scheme are similar, matching the DoF outer
bound for all antenna configurations. Comparing with the DoF region
when delayed feedback information is not available at both the transmitters
and the cognitive relay, the delayed feedback information is useful
for expanding the DoF region when $M_r < M_t+M_{\c}$
where $M_t$, $M_{\c}$, and $M_r$ are the numbers of antennas
at the transmitter, the relay, and the receiver, respectively.
If delayed feedback is not available at the cognitive relay,
the optimal DoF region is shown to be achievable except when $M_t<M_r<M_t+M_{\c}$
by the proposed retrospective interference alignment scheme. Our results quantitatively reveal the DoF gain from the cognitive relay according to antenna configurations when only delayed feedback information is available.

Finally, we compare the sum DoF of ICCR with those of broadcast channel and interference channel when only delayed CSIT is available. With the help of the cognitive relay, ICCR has an enlarged DoF region compared to MIMO interference channel.
Furthermore, as a corollary of the above-mentioned results,
lower and upper bounds are derived for the sum DoF of a 
cognitive interference channel which is also known as a 
interference channel with a cognitive transmitter.

\subsection{Paper Organization}\label{organization}
This paper is organized as follows. Section~\ref{model}
describes the system model with various types of delayed
feedback information. In Section~\ref{SISO}, focusing
on the SISO model as a special case, we propose a modified
retrospective interference alignment scheme achieving the outer bound
for SISO model under various types of delayed feedback information.
Section~\ref{MIMO_cr} derives the DoF regions for the multiple antenna scenarios.
Section~\ref{achievable_dof}  derives the achievable DoF region when delayed feedback information
is not available at the cognitive relay.
In Section~\ref{outerbound}, we derive the DoF
outer bounds with and without delayed feedback information.
Section~\ref{discussion} discusses a comparison with BC and IC under delayed CSIT
and an extension to cognitive interference channel.
Section~\ref{conclusion} concludes the paper.

\section{System Model}
\label{model}
\begin{figure}
\centering
    \includegraphics[width=0.8\columnwidth]{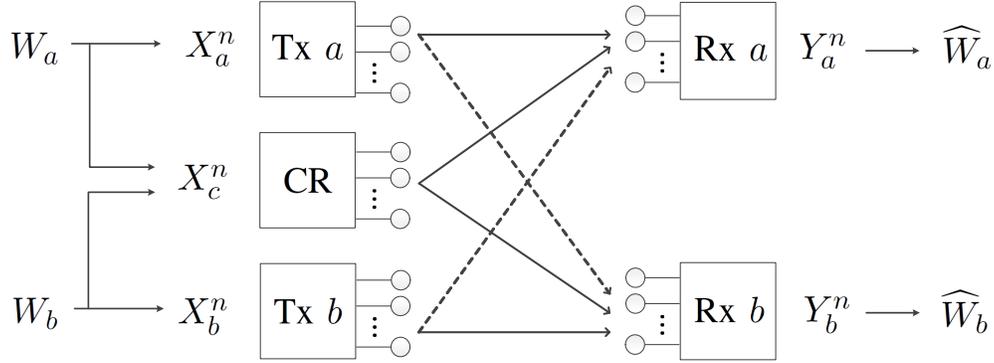}
    \caption{A 
MIMO interference channel with a cognitive relay.} \label{fig_ic_cr}
\end{figure}

This paper considers a MIMO network consisting of two transmitters
with $M_t$ antennas, two receivers with $M_r$ antennas, and a
cognitive relay with $M_{\c}$ antennas as shown in Fig.~\ref{fig_ic_cr}, where the desired and interference links are represented
by solid and dashed lines, respectively. The links experience i.i.d. Rayleigh fading.
Transmitter~$\a$ has message $W_{\a}$ intended for
Receiver~$\a$, Transmitter~$\b$ has message $W_{\b}$ intended for Receiver~$\b$, and the
cognitive relay has both $W_{\a}$ and $W_{\b}$ non-causally, where the
messages $W_{\a}$ and $W_{\b}$ are independent. Channel outputs at time
slot $t$ are
\begin{subequations}
\begin{equation}
    Y_{\a,t}=\mathbf{H}_{\a\a,t}X_{\a,t}+\mathbf{H}_{\a\b,t}X_{\b,t}+\mathbf{H}_{\a\c,t}X_{\c,t}+Z_{\a,t}, \label{eq1_II}
\end{equation}
\begin{equation}
    Y_{\b,t}=\mathbf{H}_{\b\a,t}X_{\a,t}+\mathbf{H}_{\b\b,t}X_{\b,t}+\mathbf{H}_{\b\c,t}X_{\c,t}+Z_{\b,t}, \label{eq2_II}
\end{equation}
\end{subequations}
where $Y_{j,t}=[Y_{j[1],t},\cdots,Y_{j[M_r],t}]^T
\in\mathbb{C}^{M_r\times 1},\ j\in\{\a,\b\}$, is the received signal at
Receiver~$j$, $Y_{j[\ell],t}$ is the $\ell$-th element of $Y_{j,t}$,
$X_{i,t} \in\mathbb{C}^{M_t\times 1},\ i\in \{\a,\b\}$, is
the transmitted signal from Transmitter~$i$, $X_{\c,t}
\in\mathbb{C}^{M_{\c}\times 1}$ is the transmitted signal from the
cognitive relay, $\mathbf{H}_{ji,t}\in\mathbb{C}^{M_r\times M_t}$ is
the time varying channel matrix from Transmitter~$i$ to Receiver~$j$,
$\mathbf{H}_{j\c,t}\in\mathbb{C}^{M_r\times M_{\c}}$ is time varying
channel matrix from the cognitive relay to Receiver~$j$, and $Z_{j,t}$\footnote{This noise terms can be ignored since this paper considers a high signal-to-noise (SNR) model.} is
an i.i.d. circular symmetric complex Gaussian noise,
$\mathcal{CN}(0,I_{M_r})$, at Receiver~${j}$. We assume that all channel
coefficients are i.i.d. circular symmetric complex Gaussian random
variables with zero mean and unit variance, $\mathcal{CN}(0,1)$.

In this paper, we assume perfect CSIR. Certain feedback information is available at the transmitters and cognitive relay with delay, represented by the following four cases, where $i\in \{\a,\b\}$ and $t$ is the time index:
\begin{enumerate}
\item Delayed CSIT: $X_{i,t}=f_{i,t}(W_{i},\mathcal{H}^{t-1})$, $X_{\c,t}=f_{\c,t}(W_{\a},W_{\b},\mathcal{H}^{t-1})$
\item Delayed output feedback: $X_{i,t}=f_{i,t}(W_{i},{Y}_i^{t-1})$, $X_{\c,t}=f_{\c,t}(W_{\a},W_{\b},{Y}_{\a}^{t-1},{Y}_{\b}^{t-1})$
\item Delayed Shannon feedback: $X_{i,t}=f_{i,t}(W_{i},{Y}_i^{t-1},\mathcal{H}^{t-1})$, $X_{\c,t}=f_{\c,t}(W_{\a},W_{\b},{Y}_{\a}^{t-1},{Y}_{\b}^{t-1},\mathcal{H}^{t-1})$
\item No feedback: $X_{i,t}=f_{i,t}(W_{i}),$ $X_{\c,t}=f_{\c,t}(W_{\a},W_{\b}).$
\end{enumerate}

Each message $W_i \in\left\{1,2,\cdots,2^{nR_i(P)}\right\}$ is
uniformly distributed, $f_{i,t}$ and $f_{c,t}$ are, respectively, encoding functions at Transmitter $i$ and the cognitive relay for channel use $t$ and $\mathcal{H}_t$ is the set of all
channel matrices at time index $t$, i.e.,
\begin{align*}
\mathcal{H}_t&\triangleq\{\mathbf{H}_{\a\a,t}, \mathbf{H}_{\a\b,t},
\mathbf{H}_{\a\c,t}, \mathbf{H}_{\b\a,t},
\mathbf{H}_{\b\b,t},\mathbf{H}_{\b\c,t}\},\\
\mathcal{H}^{t}&\triangleq\left\{\mathcal{H}_1,\mathcal{H}_2,\cdots,\mathcal{H}_t\right\}.
\end{align*}

$X_{i,t}$ and $X_{\c,t}$ should satisfy the power constraint
$\mathbb{E}\big[ ||X_{i,t}||^2\big]\leq P$ and $\mathbb{E}\big[||X_{\c,t}||^2\big]\leq P$, respectively
where $i\in \{\a,\b\}$.

Receiver $i$ decodes the message
from the received signal with a decoding function $g_{i}$ such that
$\widehat{W_i}=g_{i}(Y_i^n,\mathcal{H}^n)$.

A rate pair $(R_{\a}(P),R_{\b}(P))$ is achievable if there exists a
sequence of codes $\left(2^{nR_{\a}(P)},2^{nR_{\b}(P)},n\right)$ whose
average probability of error goes to zero as $n\rightarrow\infty$. The
capacity region $\mathcal{C}(P)$ is defined as the set of all
achievable rate pairs $(R_{\a}(P),R_{\b}(P))$, and the DoF region can
be defined from the capacity region as
\begin{align}
    \mathcal{D}=\Big\{(d_{\a},d_{\b})\in\mathbb{R}_{+}^2|&\ \forall (w_{\a}, w_{\b}) \in \mathbb{R}_{+}^2,\nonumber\\
     &\ \  w_{\a} d_{\a}+w_{\b} d_{\b} \leq \limsup_{P \to \infty}\frac{1}{\mathrm{log_2}P}\big[\sup_{(R_{\a}(P),R_{\b}(P))\in\mathcal{C}(P)}{w_{\a}R_{\a}(P)+w_{\a}R_{\b}(P)} \big]\Big\}. \nonumber
\end{align}

The element $\ell$ of the received vector $Y_{i,t}$ at time index
$t$ is denoted as $Y_{i[\ell],t}$. In a similar manner, a subset of
elements from this vector is denoted as follows:
\[
Y_{i[\ell_1:\ell_2],t}\triangleq\{Y_{i[\ell_1],t},Y_{i[\ell_1+1],t},\cdots,Y_{i[\ell_2],t}\}
\]
In the same manner, we define a sequence of vectors over all (causal) time that select only a subset of the antennas:
\[
Y_{i[\ell_1:\ell_2]}^t\triangleq\{Y_{i[\ell_1]}^t,Y_{i[\ell_1+1]}^t,\cdots,Y_{i[\ell_2]}^t\}
\]
In the special case where only one antenna is selected across time we have
$Y_{i[\ell]}^t=\{Y_{i[\ell],1},Y_{i[\ell],2},\cdots,Y_{i[\ell],t}\}$.

$g(x)=o(f(x))$ denotes that functions $g(\cdot),f(\cdot)$ have the following tail characteristic:
$\lim_{x\rightarrow \infty}\frac{g(x)}{f(x)}=0$. Several specialized
notations are shown in Table~\ref{table_notation} that distinguish the DoF regions under various
conditions.

\begin{table}
\caption{DoF Notations for 
 Interference Channel with Cognitive Relay}
\label{table_notation}
\begin{center}
\begin{tabular}{|c |l |}
\hline
$\mathcal{D}_{\mathrm{no}}$& DoF region 
with no feedback \\ \hline
$\mathcal{D}_{\mathrm{CSI}}$ & DoF region 
 with delayed CSIT \\ \hline
$\mathcal{D}_{\mathrm{output}}$& DoF region 
 with delayed output feedback \\ \hline
$\mathcal{D}_{\mathrm{Shannon}}$& DoF region 
 with delayed Shannon feedback \\ \hline
$\mathcal{D}_{\mathrm{perfect}}$& DoF region 
 with perfect CSIT\\ \hline
$\mathcal{D}'_{\mathrm{delay \backslash CR}}$ & Achievable DoF region 
 with delayed feedback unavailable at CR \\ \hline
 $\mathcal{\bar{D}}_\mathrm{\!\!~delay}$& DoF outer bound 
 with delayed feedback\\  \hline
$\mathcal{\bar{D}}_{\mathrm{no}}$& DoF outer bound 
with no feedback \\ \hline
\end{tabular}
\end{center}
\end{table}

\section{
SISO DoF with Delayed Feedback}\label{SISO}

This section focuses on the SISO special case, i.e.,
$M_t=M_r=M_{\c}=1$. We propose a modified retrospective interference alignment scheme
achieving the DoF outer bound of the 
 Gaussian i.i.d.\ fading
SISO interference channel with a cognitive relay. This is done on
the one hand when any of the three kinds of feedback information is
available, and on the other hand when no feedback is available.
Each receiver is assumed to have perfect CSI. 

\subsection{Delayed CSIT}
\label{delayed_CSIT}
\begin{figure}
\centering
    \includegraphics[width=0.9\columnwidth]{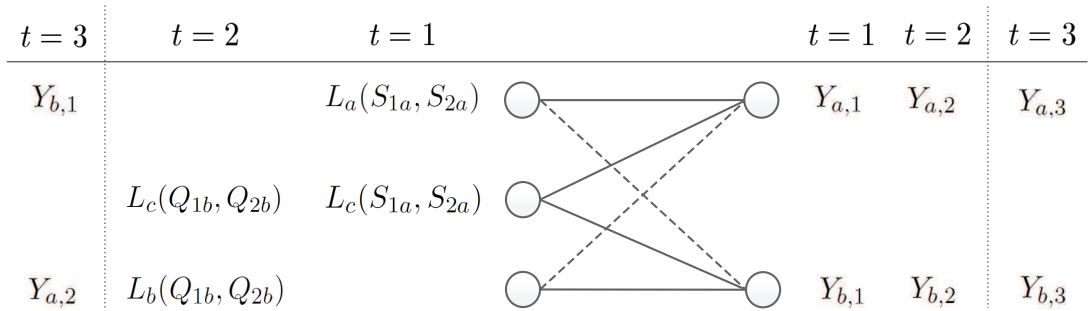}
    \caption{Achievable scheme for the delayed CSIT case,  $L(x,y)$ is a random linear combination of $x$ and $y$.} \label{fig_scheme}
\end{figure}

We now assume the transmitters and cognitive relay have perfect knowledge of all
channel gains after one time slot delay.
\begin{theorem} \label{theorem:dof_siso_csit}

The DoF region of the 
 SISO ICCR with delayed CSIT $\mathcal{D}_\mathrm{CSI}$ is
    \begin{align}
        \mathcal{D}_\mathrm{CSI}= \bigg\{(d_{\a},d_{\b})\in\mathbb{R}_{+}^2\ : \ & d_{\a}+\frac{d_{\b}}{2} \leq 1,\ \frac{d_{\a}}{2}+d_{\b} \leq 1 \bigg\} \label{dof_delayed}
    \end{align}

    where $M_t=M_r=M_{\c}=1$.
\end{theorem}
\begin{IEEEproof}
The outer bound of DoF region given in (\ref{dof_delayed}) will
be derived in Theorem~\ref{theorem:dof_mimo_outer} of Section~\ref{outerbound}.
We propose a coding scheme that achieves a
$(d_{\a},d_{\b})=(\frac{2}{3},\frac{2}{3})$ DoF pair almost surely. The
coding scheme is a modified version of the retrospective interference alignment in~\cite{MAT2012} for our channel model. The DoF tuple achieved by this
scheme is a point on the DoF region as shown in Fig.~\ref{fig_dof}. Then, we can also achieve the entire DoF region using
time sharing.

\begin{figure}
\centering
    \includegraphics[width=0.6\columnwidth]{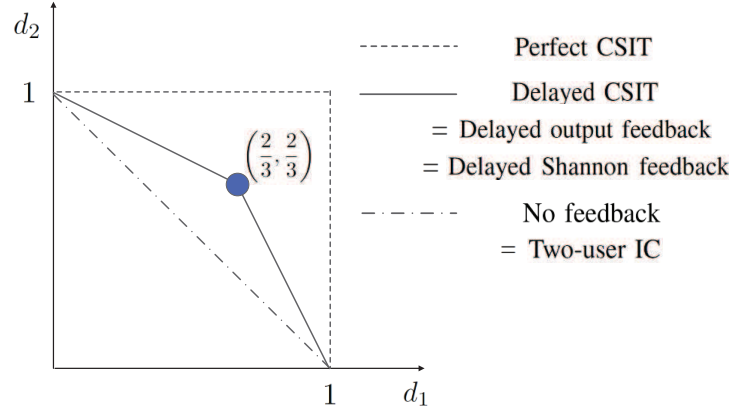}
    \caption{The DoF region of the 
SISO Gaussian interference channel
    with and without a cognitive relay.} \label{fig_dof}
\end{figure}

Now, we show that the $(\frac{2}{3},\frac{2}{3})$ DoF pair is
achievable under delayed CSIT.
Time slots are partitioned into groups of three, and each transmitter
sends two symbols during the 3 time slots, thus DoF of 2/3 is achieved
per user. The transmit symbols of Transmitter~$\a$ are denoted as
$S_{1\a}$ and $S_{2\a}$, and the transmit symbols for Transmitter~$\b$
are $Q_{1\b}$ and $Q_{2\b}$. The transmission mechanism is as follows:
in the first time slot Transmitter~$\a$ and the cognitive relay
transmit (different) random linear combinations of $S_{1\a},S_{2\a}$,
while Transmitter~$\b$ is silent. Neglecting the noise terms, the
received signals are:
\begin{subequations}
\begin{equation}
    Y_{\a,1}\!=\!H_{\a\a,1}(u_{1\a,1}S_{1\a}\!+\!u_{2\a,1} S_{2\a})\!\!+\!H_{\a\c,1}(v_{1\c,1} S_{1\a}\!+\!v_{2\c,1} S_{2\a}), \label{eq1_VI}
\end{equation}
\begin{equation}
    Y_{\b,1}\!=\!H_{\b\a,1}(u_{1\a,1}S_{1\a}\!+\!u_{2\a,1} S_{2\a})\!\!+\!H_{\b\c,1}(v_{1\c,1} S_{1\a}\!+\!v_{2\c,1} S_{2\a}). \label{eq2_VI}
\end{equation}
\end{subequations}
All precoding variables are chosen so that power constraints are
satisfied, but so that $\frac{u_{1\a,1}}{u_{2\a,1}} \neq
\frac{v_{1\c,1}}{v_{2\c,1}}$. In time slot 2, a similar action takes
place, except Transmitter~$\b$ and the cognitive relay transmit and
Transmitter~$\a$ is silent.
\begin{subequations}
\begin{equation}
    Y_{\a,2}\!=\!H_{\a\b,2}(u_{1\b,2}Q_{1\b}\!+\!u_{2\b,2} Q_{2\b})\!\!+\!H_{\a\c,2}(v_{1\c,2} Q_{1\b}\!+\!v_{2\c,2} Q_{2\b}),\label{eq3_VI}
\end{equation}
\begin{equation}
    Y_{\b,2}\!=\!H_{\b\b,2}(u_{1\b,2}Q_{1\b}\!+\!u_{2\b,2} Q_{2\b})\!\!+\!H_{\b\c,2}(v_{1\c,2} Q_{1\b}\!+\!v_{2\c,2} Q_{2\b}).\label{eq4_VI}
\end{equation}
\end{subequations}
where similar conditions on the precoding variables are imposed.
Finally, in time slot 3, Transmitter~$\a$ and Transmitter~$\b$ transmit but the relay is
silent. Using delayed CSIT, the transmitters respectively transmit
the received signal at their non-intended receiver during the
initial transmission, appropriately scaled to account for power
constraints.
\begin{subequations}
\begin{equation}
    Y_{\a,3}=H_{\a\a,3} (p_1Y_{\b,1}) + H_{\a\b,3} (p_2Y_{\a,2}), \label{eq5_VI}
\end{equation}
\begin{equation}
    Y_{\b,3}=H_{\b\a,3} (p_1Y_{\b,1}) + H_{\b\b,3} (p_2Y_{\a,2}). \label{eq6_VI}
\end{equation}
\end{subequations}

Subtracting $H_{\a\b,3}(p_2Y_{\a,2})$ from $Y_{\a,3}$ with the received
signal at time index $t=2$, Receiver~$\a$ can obtain the interference-free
signal $Y_{\b,1}$ as
\begin{align}
    Y_{\b,1}= \frac{Y_{\a,3}-H_{\a\b,3}(p_2Y_{\a,2})}{H_{\a\a,3}\!\ p_1}. \nonumber
\end{align}
The signaling scheme is depicted in Fig.~\ref{fig_scheme}. We can
readily know that $Y_{\a,1}$ and $Y_{\b,1}$ are almost surely
linearly independent since channel gains are independently drawn
from the same continuous distribution and $u_{j\a,1}$ and
$v_{j\a,1}, j\in\{1,2\}$, are also random and independent. Thus,
Receiver~$\a$ has two independent equations given by linear
combinations of two variables $S_{1\a}$ and $S_{2\a}$ so that it can
decode two symbols. Similarly, since Receiver~$\b$ can obtain
$Y_{\a,2}$, Receiver~$\b$ also has two independent equations
$Y_{\a,2}$ and $Y_{\b,2}$ of two variables intended for
Receiver~$\b$, and hence Receiver~$\b$ can decode two symbols
$Q_{1\b}$ and $Q_{2\b}$. Consequently, at the end of transmission,
each receiver can achieve $\frac{2}{3}$ DoF (i.e., two symbols over
3 time slots) almost surely. In other words, the sum DoF is
$\frac{4}{3}$.
\end{IEEEproof}\vspace{0.1in}

\begin{remark}
 The DoF region under perfect CSIT at the transmitters and cognitive relay
is~\cite{SVJS2008}
\begin{align}
        \mathcal{D}_{\mathrm{perfect}}= \Big\{(d_{\a},d_{\b})\in\mathbb{R}_{+}^2\ : \ & d_{\a}\leq 1,\ d_{\b} \leq 1 \Big\} \label{dof_perfect}
\end{align}
which is shown in Fig.~\ref{fig_dof} as a reference. With perfect
instantaneous CSI at the transmitters and cognitive relay, sum DoF is $2$ almost surely, which is as if
receivers are free from interference. The DoF achieving strategy is
interference pre-cancelation via the relay's non-causal knowledge of
the messages. On the other hand, the 
 Gaussian SISO
interference channel without cognitive relay has sum DoF of $1$
regardless of whether transmitters have CSI.
\end{remark}
\begin{remark}
Theorem~\ref{theorem:dof_siso_csit} indicates that a cognitive relay can
increase DoF even with delayed CSIT although the amount of DoF
increased by a cognitive relay is limited compared to the case of
perfect CSIT; the 
 SISO ICCR with delayed CSIT has total
$\frac{4}{3}$ DoF at most.
\end{remark}\vspace{0.2in}

\subsection{Delayed Output Feedback}
\label{output}

Each receiver feeds its output back to its transmitter so that each
transmitter knows only the output of the intended receiver after one
time slot delay. The cognitive relay also has the
output feedback from both the receivers after one time slot delay.

\begin{figure}
\centering
    \includegraphics[width=0.9\columnwidth]{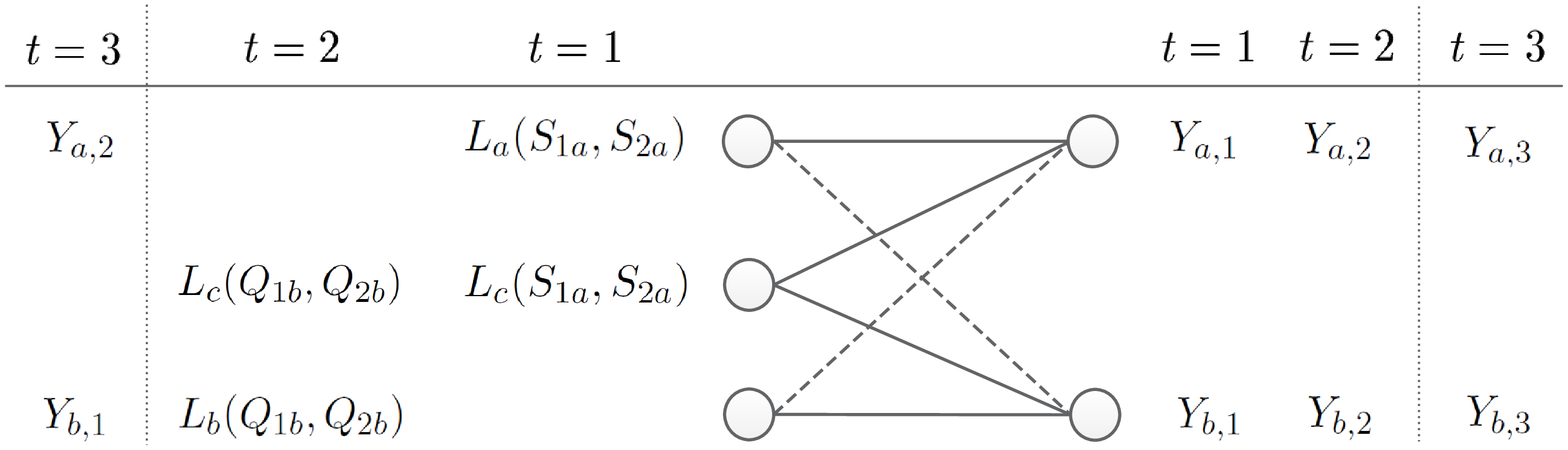}
    \caption{Achievable scheme for the delayed output feedback case,  $L(x,y)$ is a random linear combination of $x$ and $y$.} \label{fig_scheme_output}
\end{figure}

\begin{theorem}\label{theorem:dof_siso_output}
The DoF region of the 
 SISO ICCR with delayed output feedback is
\begin{align}
    \mathcal{D}_\mathrm{output}=\mathcal{D}_\mathrm{CSI}. \label{dof_output}
\end{align}

where $M_t=M_r=M_{\c}=1$.
\end{theorem}
\begin{IEEEproof}
The outer bound is determined by $d_{\a}+\frac{d_{\b}}{2} \leq
 1,\ \frac{d_{\a}}{2}+d_{\b} \leq 1$ that will be derived in Theorem
~\ref{theorem:dof_mimo_outer} of Section~\ref{outerbound}.
 We propose a scheme that achieves
 $(d_{\a},d_{\b})=(\frac{2}{3},\frac{2}{3})$ DoF pair almost surely, a
 point that is on the outer bound, and then all other points on the
 outer bound are achieved via time sharing.

Similar to the delayed CSIT case, the achievable scheme needs $3$ time
slots. At time slot 1 and 2, the signaling follows Section~\ref{delayed_CSIT}. In time slot 3, however, a different signaling
is used where the transmitter utilizes the output feedback from the
receiver instead of constructing a linear combination of previous
symbols based on delayed CSI. Each transmitter transmits the output
fed back from the intended receiver, appropriately scaled to satisfy
the power constraints.
\begin{subequations}
\begin{equation}
    Y_{\a,3}=H_{\a\a,3} (p_1Y_{\a,2}) + H_{\a\b,3} (p_2Y_{\b,1}), \label{eq7_VI}
\end{equation}
\begin{equation}
    Y_{\b,3}=H_{\b\a,3} (p_1Y_{\a,2}) + H_{\b\b,3} (p_2Y_{\b,1}), \label{eq8_VI}
\end{equation}
\end{subequations}
Subtracting $H_{\a\a,3}(p_1Y_{\a,2})$ from $Y_{\a,3}$ with the received
signal at time index $t=2$, Receiver~$\a$ can obtain the interference-free
signal $Y_{\b,1}$ as
\begin{align}
    Y_{\b,1}= \frac{Y_{\a,3}-H_{\a\a,3}(p_1Y_{\a,2})}{H_{\a\b,3}\!\ p_2}. \nonumber
\end{align}
The signaling scheme is shown in Fig.~\ref{fig_scheme_output}. Because Receiver~$\a$ almost surely has two linearly independent equations
$Y_{\a,1}$ and $Y_{\b,1}$ that are linear combinations of two symbols
$S_{1\a}$ and $S_{2\a}$, Receiver~$\a$ is able to decode the two symbols.
Similarly, because Receiver~$\b$ can obtain $Y_{\a,2}$ and almost surely has
two linearly independent equations $Y_{\a,2}$ and $Y_{\b,2}$ of two
symbols, Receiver~$\b$ can decode two symbols $Q_{1\b}$ and $Q_{2\b}$. As a
result, each receiver can achieve $\frac{2}{3}$ DoF almost surely,
and we can achieve $\frac{4}{3}$ sum DoF.
\end{IEEEproof}

\subsection{Delayed Shannon Feedback}
Shannon feedback refers to a strictly causal feedback
that gives each transmitter all the channel state information as
well as the received value at the intended receiver.
The cognitive relay has also the delayed Shannon feedback from the receivers after one time slot delay.
\begin{theorem}\label{theorem:dof_siso_shannon}
The DoF region of the 
 SISO ICCR with delayed Shannon feedback is
\begin{align}
   \mathcal{D}_\mathrm{Shannon}= \mathcal{D}_\mathrm{CSI}.  \label{dof_shannon}
\end{align}
\end{theorem}
\begin{IEEEproof}
The outer bound for delayed Shannon feedback is the same as that
for delayed CSIT feedback or delayed output feedback. The outer bound is presented in
Theorem~\ref{theorem:dof_mimo_outer} of Section~\ref{outerbound}.
The outer bound can be achieved for the cases of delayed CSIT
and delayed output feedback from Theorem~\ref{theorem:dof_siso_csit} and
Theorem~\ref{theorem:dof_siso_output}. Therefore, the outer bound is
also achievable because we can use both the delayed CSIT and the
output feedback information.
\end{IEEEproof}
\begin{remark}
For the SISO case, the proposed scheme does not entail any delayed feedback information at the cognitive relay. Therefore,
the optimal DoF region of the SISO ICCR can be obtained even if the cognitive relay does not have delayed feedback.
\end{remark}
\subsection{No Feedback}\label{no}

\begin{corollary} \label{corollary:dof_siso_no}
The DoF region for the SISO ICCR with no feedback is
    \begin{align}
        \mathcal{D}_{\mathrm{no}}  = \Big\{(d_{\a},d_{\b})\in\mathbb{R}_{+}^2\ : \ & d_{\a}+d_{\b}\leq 1 \Big\}.   \label{dof_no}
    \end{align}
\end{corollary}
\begin{IEEEproof}
The DoF outer bound is $d_{\a}+d_{\b}\leq 1$ that will be proved in
Corollary~\ref{corollary:dof_mimo_outer_no} of Section~\ref{outerbound}.
The outer bound is achievable via time
division multiplexing (TDM) when the transmitters and cognitive relay
do not have any feedback information.
\end{IEEEproof}
The result is true irrespective of the number of transmit or receive
antennas. In section~\ref{MIMO_no}, we will show that TDM is also
DoF optimal for the MIMO case.
\begin{remark}
The DoF region in Corollary~\ref{corollary:dof_siso_no} is the same as
that of the 
 SISO interference channel without a cognitive relay. This shows the
cognitive relay in the SISO case has no effect on DoF in the absence
of CSIT.
\end{remark}

\section{MIMO DoF with Delayed Feedback}\label{MIMO_cr}
This section extends the modified retrospective interference alignment scheme
and applies it to multi-antenna nodes. We derive achievable DoF region for
four types of feedback information (including no feedback). Each transmitter has $M_t$
antennas, each receiver has $M_r$ antennas, and the cognitive relay
has $M_{\c}$ antennas. We continue to assume perfect CSIR.

\subsection{Delayed CSIT}\label{MIMO_csit}
The transmitters and cognitive relay have perfect knowledge of all channel information after
one time slot delay The analysis is divided into five categories according to antenna configuration (see
Table~\ref{table1}).

\begin{table}
\caption{The five conditions according to antenna configurations }
\label{table1}
\begin{center}
\begin{tabular}{|c |c |c|c|c|}
\hline
 \multirow{2}{*}{$M_t+M_{\c}\leq M_r~$} &\multicolumn{2}{c|}{$\frac{M_t+M_{\c}}{2}\leq M_r<M_t+M_{\c}$} &\multicolumn{2}{c|}{$M_r <\frac{M_t+M_{\c}}{2}$}\\ \cline{2-5}
 &$M_r >M_t$ &$~M_r\leq M_t$ &$M_r >M_t$ &$~M_r\leq M_t$\\  \hline\hline
Condition I & Condition II & Condition III & ~Condition IV & Condition V \\ \hline
\end{tabular}
\end{center}
\end{table}

\begin{theorem} \label{theorem:dof_mimo_cr}
The DoF region of the MIMO ICCR with delayed CSIT is
\begin{align}
   \mathcal{D}_\mathrm{CSI} = \bigg\{&(d_{\a},d_{\b})\in\mathbb{R}_{+}^2\ :\ \;  d_{\a}\leq \min(M_r,M_t+M_{\c}),\;d_{\b}\leq \min(M_r,M_t+M_{\c}), \nonumber\\
   & \frac{d_{\a}}{\min(M_r,M_t+M_{\c})}+\frac{d_{\b}}{\min(2M_r,M_t+M_{\c})}\leq \frac{\min(M_r,2M_t+M_{\c})}{\min(M_r,M_t+M_{\c})},\nonumber\\
   & \frac{d_{\a}}{\min(2M_r,M_t+M_{\c})}+\frac{d_{\b}}{\min(M_r,M_t+M_{\c})}\leq \frac{\min(M_r,2M_t+M_{\c})}{\min(M_r,M_t+M_{\c})} \bigg\} \nonumber
\end{align}
where $M_t, ~M_r$, and $M_{\c}$ are the number of antennas at the
transmitter, the receiver and the cognitive relay, respectively.
\end{theorem}
\begin{IEEEproof}
We show the achievable DoF region according to the classified
conditions, and compare the achievable DoF region with the DoF outer
bound that will be derived in Theorem~\ref{theorem:dof_mimo_outer}
of Section~\ref{outerbound}. The delayed CSIT is
not used in the achievable scheme for Condition I, but we
exploit delayed CSIT for Condition II, III, IV and V.

\begin{enumerate}
\item Condition I: $M_t+M_{\c}\leq M_r$

In this case, the DoF outer bound with delayed
CSIT is constructed from $d_{\a}\leq M_t+M_{\c},\ d_{\b}\leq M_t+M_{\c},$
and $d_{\a}+d_{\b}\leq \min(M_r, 2M_t+M_{\c})$. Using a similar approach of decomposing an interference channel into multiple access channel (MAC) \cite{C1978}, we decompose the ICCR into MACs.
The two users and the cognitive
relay each use their own codewords. Since each receiver can decode
the maximum of $\min(M_r, 2M_t+M_{\c})$ signals, the two transmitters and cognitive
relay send total $\min(M_r, 2M_t+M_{\c})$ messages, then each receiver can decode
all signals. Consequently, the optimal DoF region is obtained and the total sum DoF becomes $\min(M_r, 2M_t+M_{\c})$.

\item Condition II: $\frac{M_t+M_{\c}}{2}\leq M_r<M_t+M_{\c}$ and $M_r >M_t$

A DoF outer bound with delayed CSIT for this case is given by
$\frac{d_{\a}}{M_t+M_{\c}}+\frac{d_{\b}}{M_r}\leq 1$ and
$\frac{d_{\a}}{M_r}+\frac{d_{\b}}{M_t+M_{\c}}\leq 1$.
We can show that the DoF pair
$\Big(\frac{(M_t+M_{\c})M_r}{M_r+M_t+M_{\c}},\frac{(M_t+M_{\c})M_r}{{M_r}+M_t+M_{\c}}\Big)$
 is achievable, which lies on the intersection of
$\frac{d_{\a}}{M_t+M_{\c}}+\frac{d_{\b}}{M_r}\leq 1$ and
$\frac{d_{\a}}{M_r}+\frac{d_{\b}}{M_t+M_{\c}}\leq 1$ on the DoF
outer bound.
First, for each time slot
$t\in \{1,\cdots,M_r\}$, Transmitter~$\a$ sends random linear
combinations of $M_t+M_{\c}$ independent symbols, and the cognitive
relay sends different random linear combinations of the $M_t+M_{\c}$
independent symbols. The received signals at time index $t\in
\{1,\cdots,M_r\}$ can be represented as
\begin{subequations}
\begin{equation}
    Y_{\a,t}=\mathbf{H}_{\a\a,t}\mathbf{U}_{\a,t}S_{\a,t}+\mathbf{H}_{1\c,t}\mathbf{V}_{\c,t}S_{\a,t}, \label{eq1_III}
\end{equation}
\begin{equation}
    Y_{\b,t}=\mathbf{H}_{\b\a,t}\mathbf{U}_{\a,t}S_{\a,t}+\mathbf{H}_{2\c,t}\mathbf{V}_{\c,t}S_{\a,t}, \label{eq2_III}
\end{equation}
\end{subequations}
where $\mathbf{U}_{\a,t}$ is a randomly chosen $M_t\times (M_t+M_{\c})$
matrix with rank $M_t$, $\mathbf{V}_{\c,t}$ is a randomly chosen
$M_{\c}\times (M_t+M_{\c})$ matrix with rank $M_{\c}$, $S_{\a,t}$ is
an $(M_t+M_{\c}) \times 1$ symbol vector for Receiver~$\a$ at time index
$t$, the transmissions are appropriately scaled to satisfy the power
constraint, and noise terms are omitted since noise does not affect
DoF. Because Receiver~$\a$ obtains $M_r$ linear combinations of the desired
$M_t+M_{\c}$ variables at each time index, Receiver~$\a$ has total $M_r^2$
independent linear equations of the $(M_t+M_{\c})M_r$ desired
symbols during $M_r$ time slots.

Then in each time slot $t\in
\{M_r+1,\cdots,2M_r\}$, Transmitter~$\b$ and the cognitive relay send
different random linear combinations of $M_t+M_{\c}$ symbols
intended for Receiver~$\b$. The received signals at time index $t\in
\{M_r+1,\cdots,2M_r\}$ are
\begin{subequations}
\begin{equation}
    Y_{\a,t}=\mathbf{H}_{\a\b,t}\mathbf{U}_{\b,t}Q_{\b,t}+\mathbf{H}_{1\c,t}\mathbf{V}_{\c,t}Q_{\b,t}, \label{eq3_III}
\end{equation}
\begin{equation}
    Y_{\b,t}=\mathbf{H}_{\b\b,t}\mathbf{U}_{\b,t}Q_{\b,t}+\mathbf{H}_{2\c,t}\mathbf{V}_{\c,t}Q_{\b,t}, \label{eq4_III}
\end{equation}
\end{subequations}
where $\mathbf{U}_{\b,t}$ is a randomly chosen $M_t\times (M_t+M_{\c})$
matrix with rank $M_t$, $\mathbf{V}_{\c,t}$ is a randomly chosen
$M_{\c}\times (M_t+M_{\c})$ matrix with rank $M_{\c}$, $Q_{\b,t}$ is
an $(M_t+M_{\c})\times 1$ symbol vector for Receiver~$\b$ at time index
$t$, all coefficients are appropriately selected to satisfy the
power constraint, and noise terms are omitted. Receiver~$\b$ obtains total
$M_r^2$ independent linear combinations of the $(M_t+M_{\c})M_r$
desired variables during $M_r$ time slots.

At time index
$t\in\{2M_r+1,...,M_t+M_{\c}+M_r\}$, Transmitter~$\a$,
Transmitter~$\b$, and the cognitive relay transmit
$X_{\a,t}=[Y_{\b[t-2M_r],1},...,Y_{\b[t-2M_r],M_t}]^T$,
$X_{\b,t}=[Y_{\a[t-2M_r],M_r+1},...,Y_{\a[t-2M_r],M_r+M_t}]^T$ and
$X_{\c,t}=[Y_{\b[t-2M_r],M_t+1}+Y_{\a[t-2M_r],M_r+M_t+1},...,Y_{\b[t-2M_r],M_r}+Y_{\a[t-2M_r],2M_r},0,...,0]^T$,
respectively, using delayed CSI. Note that the cognitive relay
transmits $X_{\c,t}$ using only $M_r -M_t$ antennas.
The transmissions are appropriately scaled to satisfy the power
constraint. The received signals at $t\in\{2M_r+1,\cdots,
M_t+M_{\c}+M_r\}$ are
\begin{subequations}
\begin{equation}
    Y_{\a,t}=\mathbf{H}_{\a\a,t}X_{\a,t}+\mathbf{H}_{\a\b,t}X_{\b,t}+\mathbf{H}_{\a\c,t}X_{\c,t}, \label{eq5_III}
\end{equation}
\begin{equation}
    Y_{\b,t}=\mathbf{H}_{\b\a,t}X_{\a,t}+\mathbf{H}_{\b\b,t}X_{\b,t}+\mathbf{H}_{\b\c,t}X_{\c,t}, \label{eq6_III}
\end{equation}
\end{subequations}
where noise terms are omitted. Since the interfering terms are
comprised of the past received signal in previous slots, each receiver
can eliminate the interfering terms using the received signals in
the previous time slots and obtain $(M_t+M_{\c}-M_r)M_r$ linearly
independent interference-free signals during $M_t+M_{\c}-M_r$ time
slots. Therefore, each receiver has  $(M_t+M_{\c})M_r$ linearly
independent equations involving $(M_t+M_{\c})M_r$ symbols and thus
we can obtain the DoF pair
$\Big(\frac{(M_t+M_{\c})M_r}{M_t+M_{\c}+M_r},\frac{(M_t+M_{\c})M_r}{M_t+M_{\c}+M_r}\Big)$
which is the same achievable DoF pair in Condition II. The other
points on the DoF outer bound can be also achieved via time sharing.

\item Condition III: $\frac{M_t+M_{\c}}{2}\leq M_r <M_t+M_{\c}$ and $M_r \leq M_t$

We show that the
$\Big(\frac{(M_t+M_{\c})M_r}{M_t+M_{\c}+M_r},\frac{(M_t+M_{\c})M_r}{M_t+M_{\c}+M_r}\Big)$ DoF pair is
achievable, which is an intersection point of
$\frac{d_{\a}}{M_t+M_{\c}}+\frac{d_{\b}}{M_r}\leq 1$ and
$\frac{d_{\a}}{M_r}+\frac{d_{\b}}{M_t+M_{\c}}\leq 1$ on the DoF outer bound.

First, we spend $2M_r$ time slots. At each time index $t\in
\{1,\cdots,M_r\}$, Transmitter~$\a$ sends random linear combinations of
$M_t+M_{\c}$ independent symbols, and the cognitive relay sends
different random linear combinations of the $M_t+M_{\c}$ independent
symbols. Since Receiver~$\a$ obtains $M_r$ linear combinations of the
desired $M_t+M_{\c}$ variables at each time index, Receiver~$\a$ has total
$M_r^2$ independent linear equations of the $(M_t+M_{\c})M_r$
desired symbols during $M_r$ time slots. At each time index $t\in
\{N+1,\cdots,2M_r\}$, Transmitter~$\b$ and the cognitive relay send different
random linear combinations of $M_t+M_{\c}$ symbols intended for Receiver~$\b$. Receiver~$\b$ obtains total $M_r^2$ independent linear combinations of
the $(M_t+M_{\c})M_r$ desired variables during $M_r$ time slots.

Second, we need $M_t+M_{\c}-M_r$ time slots. At time index
$t\in\{2M_r+1,\cdots,M_t+M_{\c}+M_r\}$, Transmitter~$\a$ and Transmitter~$\b$ transmit
$X_{\a,t}=[Y_{\b[t-2M_r],1},\cdots,Y_{\b[t-2M_r],M_r},0,\cdots,0]^T$
and $X_{\b,t}=[Y_{\a[t-2M_r],M_r+1},\cdots,Y_{\a[t-2M_r],2M_r},$
$0,\cdots,0]^T$, respectively, using only $M_r$ antennas. The
cognitive relay does not transmit. Since Receiver~$\a$ knows
$Y_{\a[t-2M_r],M_r+1},$ $\cdots, Y_{\a[t-2M_r],2M_r}$ and Receiver~$\b$ knows
$Y_{\b[t-2M_r],1}, \cdots, Y_{\b[t-2M_r],M_r}$ where
$t\in\{2M_r+1,\cdots,M_t+M_{\c}+M_r\}$, each receiver can eliminate
interference terms and obtain $(M_t+M_{\c}-M_r)M_r$ linearly
independent interference-free signals during $M_t+M_{\c}-M_r$ time
slots. Therefore, the receivers have total $(M_t+M_{\c})M_r$
linearly independent equations involving $(M_t+M_{\c})M_r$ symbols,
respectively. Consequently, we can obtain the
$\Big(\frac{(M_t+M_{\c})M_r}{M_t+M_{\c}+M_r},\frac{(M_t+M_{\c})M_r}{M_t+M_{\c}+M_r}\Big)$
DoF pair. The other points on the outer bound are achievable via
time sharing.

\item Condition IV: $M_r <\frac{M_t+M_{\c}}{2}$ and $M_r >M_t$

The DoF outer bound is determined by
$\frac{d_{\a}}{2M_r}+\frac{d_{\b}}{M_r}\leq 1$ and $
\frac{d_{\a}}{M_r}+\frac{d_{\b}}{2M_r}\leq 1$. We show that the
$(\frac{2M_r}{3},\frac{2M_r}{3})$ DoF pair on the DoF outer bound is
achievable. All transmissions are scaled to satisfy the power constraint.
First, we spend two time slots. At time index $t=1$,
Transmitter~$\a$ sends $M_t$ random linear combinations of $2M_r$
symbols, and the cognitive relay sends $2M_r-M_t$ different random
linear combinations of the $2M_r$ symbols.
The received signals at time index $t=1$ can be represented as
\begin{subequations}
\begin{equation}
   Y_{\a,1}=\mathbf{H}_{\a\a,1}\mathbf{U}_{\a,1}S_{\a}+\mathbf{H}_{\a\c,1}\mathbf{V}_{\c,1}S_{\a} \label{eq1_V}
\end{equation}
\begin{equation}
  Y_{\b,1}=\mathbf{H}_{\b\a,1}\mathbf{U}_{\a,1}S_{\a}+\mathbf{H}_{\b\c,1}\mathbf{V}_{\c,1}S_{\a}, \label{eq2_V}
\end{equation}
\end{subequations}
where $\mathbf{U}_{\a,1}$ is a randomly chosen $M_t\times (2M_r)$ matrix
with full rank, $\mathbf{V}_{\c,1}$ is a randomly chosen $M_{\c} \times
(2M_r)$ matrix with rank $M_r$ which includes a $(M_{\c}-M_r)\times (2M_r)$ zero matrix,
$S_{\a}$ is a $(2M_r) \times 1$ symbol vector for Receiver~$\a$, and noise
terms are omitted. Receiver~$\a$ has $M_r$ linear combinations of intended $2M_r$
variables.
Similarly, at time index
$t=2$, Transmitter~$\b$ sends $M_t$ random linear combinations of
$2M_r$ symbols intended for Receiver~$\b$, and the cognitive relay
sends different $2M_r-M_t$ random linear combinations of the $2M_r$
symbols. The received signals at time index $t=2$ are
\begin{subequations}
\begin{equation}
    Y_{\a,2}=\mathbf{H}_{\a\b,2}\mathbf{U}_{\b,2}Q_{\b}+\mathbf{H}_{\a\c,2}\mathbf{V}_{\c,2}Q_{\b}, \label{eq3_V}
\end{equation}
\begin{equation}
    Y_{\b,2}=\mathbf{H}_{\b\b,2}\mathbf{U}_{\b,2}Q_{\b}+\mathbf{H}_{\b\c,2}\mathbf{V}_{\c,2}Q_{\b}, \label{eq4_V}
\end{equation}
\end{subequations}
where $\mathbf{U}_{\b,2}$ is a randomly chosen $M_t\times (2M_r)$ matrix
with full rank, $\mathbf{V}_{\c,2}$ is a randomly chosen $M_{\c} \times
(2M_r)$ matrix with rank $M_r$ which includes a $(M_{\c}-M_r)\times
(2M_r)$ zero matrix, and $Q_{\b}$ is a $(2M_r) \times 1$ symbol
vector for Receiver~$\b$. Receiver~$\b$ obtains $M_r$ linear combinations of intended
$2M_r$ variables.

Second, we need one time slot indexed by $t=3$. Transmitter~$\a$,
Transmitter~$\b$, and the cognitive relay transmit
$X_{\a,3}=[Y_{\b[1],1},...,Y_{\b[M_t],1}]^T$,
$X_{\b,3}=[Y_{\a[1],2},...,Y_{\a[M_t],2}]^T$ and
$X_{\c,3}=[Y_{\b[M_t+1],1}+Y_{\a[M_t+1],2},...,Y_{\b[M_r],1}+Y_{\a[M_r],2},0,...,0]^T$,
respectively, using delayed CSI. At time index $t=3$, the received signals are
\begin{subequations}
\begin{equation}
    Y_{\a,3}=\mathbf{H}_{\a\a,3}X_{\a,3}+\mathbf{H}_{\a\b,3}X_{\b,3}+\mathbf{H}_{\a\c,3}X_{\c,3}, \label{eq5_V}
\end{equation}
\begin{equation}
    Y_{\b,3}=\mathbf{H}_{\b\a,3}X_{\a,3}+\mathbf{H}_{\b\b,3}X_{\b,3}+\mathbf{H}_{\b\c,3}X_{\c,3}. \label{eq6_V}
\end{equation}
\end{subequations}
Since the interference terms at each
receiver are comprised of the received signals in the previous time
slots, each receiver can obtain $M_r$ linearly independent
interference-free signals at $t=3$. Thus, each receiver has total
$2M_r$ linearly independent equations involving $2M_r$ symbols and
thus we can obtain the $(\frac{2M_r}{3},\frac{2M_r}{3})$ DoF pair
which is the same achievable DoF pair in Condition V. We can achieve
all points on the DoF outer bound via time sharing.

\item Condition V: $M_r <\frac{M_t+M_{\c}}{2}$ and $M_r \leq M_t$

In this case a DoF outer
bound is given by $\frac{d_{\a}}{2M_r}+\frac{d_{\b}}{M_r}\leq 1$ and $
\frac{d_{\a}}{M_r}+\frac{d_{\b}}{2M_r}\leq 1$. We show that the
$(\frac{2M_r}{3},\frac{2M_r}{3})$ DoF pair on the DoF outer bound is
achievable.

At time index $t=1$, if $2M_r > M_t$, Transmitter~$\a$ sends $M_t$ random linear combinations of $2M_r$ symbols with
$M_t$ transmit antennas, and the cognitive relay sends $2M_r-M_t$
different random linear combinations of the $2M_r$ symbols. If $2M_r\leq
M_t$, then Transmitter~$\a$ only transmits and the cognitive relay is silent. The
received signals at time index $t=1$ can be represented as
(\ref{eq1_V}) and (\ref{eq2_V}) where $\mathbf{U}_{\a,1}$ is a randomly
chosen $M_t\times 2M_r$ matrix with full rank, $\mathbf{V}_{\c,1}$ is a
randomly chosen $M_{\c} \times 2M_r$ matrix with rank $(2M_r-M_t)^{+}$ which
includes a $(M_{\c}-(2M_r-M_t)^{+})\times 2M_r$ zero matrix, $(x)^+=\max(x,0)$,
$S_{\a}$ is a $2M_r \times 1$ symbol vector for Receiver~$\a$, and noise
terms are omitted. Receiver~$\a$ has $M_r$ linear combinations of intended $2M_r$
variables. Similarly, at time index $t=2$, Transmitter~$\b$ sends $M_t$ random
linear combinations of $2M_r$ symbols intended for Receiver~$\b$, and the
cognitive relay sends different $2M_r-M_t$ random linear combinations of
the $2M_r$ symbols if $2M_r > M_t$. If $2M_r\leq M_t$, then Transmitter~$\b$ only
transmits and the cognitive relay becomes silent. The received
signals at time index $t=2$ are the same as (\ref{eq3_V}) and
(\ref{eq4_V}) where $\mathbf{U}_{\b,2}$ is a randomly chosen $M_t\times
2M_r$ matrix with full rank, $\mathbf{V}_{\c,2}$ is a randomly chosen
$M_{\c} \times 2M_r$ random matrix with rank $(2M_r-M_t)^{+}$ which includes a
$(M_{\c}-(2M_r-M)^{+})\times 2M_r$ zero matrix, and $Q_{\b}$ is a $2M_r
\times 1$ symbol vector for Receiver~$\b$. Receiver~$\b$ obtains $M_r$ linear
combinations of the intended $2M_r$ variables.

At time index $t=3$, Transmitter~$\a$ and Transmitter~$\b$ only transmit
$X_{\a,3}=[Y_{\b[1],1},\cdots,Y_{\b[M_r],1},0,$  $\cdots,0]^T$ and
$X_{\b,3}=[Y_{\a[1],2},\cdots,Y_{\a[M_r],2},0,\cdots,0]^T$,
respectively, using $M_r$ antennas, and the cognitive relay does not
transmit. Since the interference signals at each receiver at $t=3$
are the received signals in previous time slots, each receiver can
obtain $M_r$ linearly independent interference-free signals at
$t=3$. Thus, each receiver has total $2M_r$ linearly independent
equations involving $2M_r$ symbols and hence we can obtain the
$(\frac{2M_r}{3},\frac{2M_r}{3})$ DoF pair on the DoF outer bound.
The other points on the outer bound are achievable via time sharing.
\end{enumerate}
\end{IEEEproof}

\begin{figure}
\centering
    \includegraphics[width=0.6\columnwidth]{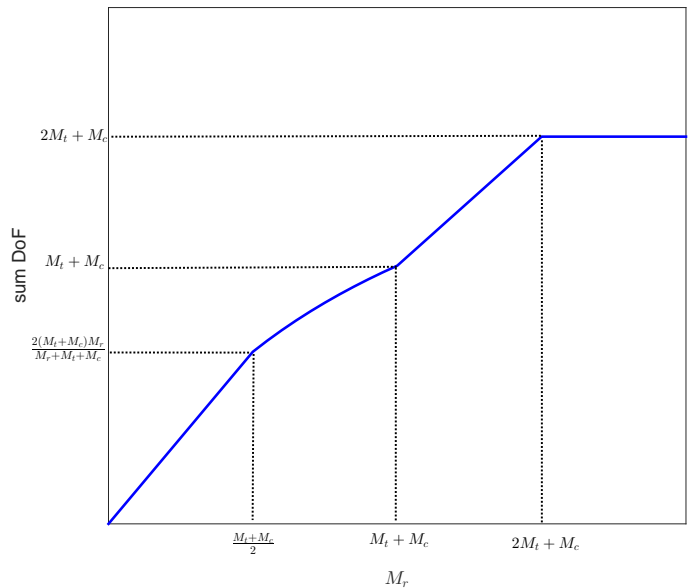}
    \caption{The sum DoF of the 
MIMO Gaussian ICCR with delayed CSIT for fixed $M_{t}$ and $M_{\c}$} \label{fig_dof_opt}
\end{figure}

\begin{remark}
Fig. \ref{fig_dof_opt} shows the result of Theorem \ref{theorem:dof_mimo_cr}
in terms of sum DoF for fixed $M_{t}$ and $M_{\c}$.
For Condition I, III, and V, the cognitive relay does
not utilize delayed feedback information. In other words, except when
$M_t<M_r<M_t+M_{\c}$, the optimal DoF region can be obtained regardless
of the availability of delayed feedback information at the cognitive relay.
This optimal DoF region will be again addressed in Section \ref{achievable_dof}.

\end{remark}

\subsection{Delayed Output Feedback}\label{MIMO_output}

In this case each transmitter knows the output of the intended
receiver after one time slot delay, and the cognitive relay has the
output feedback from both receivers after one time slot delay. The DoF
region is characterized as follows.
\begin{corollary} \label{corollary:dof_mimo_output}
The DoF region of the MIMO ICCR with delayed
output feedback is
\begin{align}
    \mathcal{D}_{\mathrm{output}}= \mathcal{D}_{\mathrm{CSI}}.
\end{align}
\end{corollary}
\begin{IEEEproof}
For Condition I, the outer bound is achievable similar to the
case of delayed CSIT, since the related scheme does not exploit any
delayed feedback information. For Conditions II, III, IV, and V, the
achievable scheme is an extension of the SISO scheme
using delayed output feedback, which has three parts. First, Transmitter~$\a$ and cognitive relay
send messages of Receiver~$\a$. Second, Transmitter~$\b$ and cognitive relay transmit
messages for Receiver~$\b$ during different time slots as in the scheme for
delayed CSIT. Third, the transmitters and cognitive relay transmit the outputs fed back
from the receivers in previous time slots, instead of transmitting
linear combinations of the past symbols with delayed CSI. Then,
similar to the delayed CSIT case, the receivers can eliminate
interference terms since the interference signals are already known at
each receiver. Thus, the DoF region with delayed output feedback
is the same as that of the delayed CSIT case.
\end{IEEEproof}
\subsection{Delayed Shannon Feedback}
\begin{corollary}\label{corollary:dof_mimo_shannon}
The DoF region of the MIMO ICCR with delayed Shannon feedback is
\begin{align}
    \mathcal{D}_{\mathrm{Shannon}}= \mathcal{D}_{\mathrm{CSI}}.
\end{align}
\end{corollary}
\begin{IEEEproof}
Since the DoF outer bounds with delayed Shannon feedback are identical to those with delayed CSIT or delayed output feedback, the same optimal DoF region can be obtained with the scheme utilizing delayed CSIT or output feedback information. The outer bound will be proved in Section~\ref{outerbound}.
\end{IEEEproof}

\subsection {No Feedback}\label{MIMO_no}

\begin{corollary} \label{theorem:dof_mimo_no}
The DoF region of the MIMO ICCR with no feedback is
    \begin{align}
        \mathcal{D}_{\mathrm{no}}=\Big\{(d_{\a},d_{\b})\in\mathbb{R}_{+}^2\ : \ & \ d_{\a}\leq \min(M_t+M_{\c},M_r),
        \; d_{\b}\leq \min(M_t+M_{\c},M_r),\nonumber\\
        & d_{\a}+d_{\b}\leq \min(2M_t+M_{\c},M_r) \Big\} \label{dof_mimo_no}
    \end{align}
where $M_t, ~M_r$, and $M_{\c}$ are the numbers of antennas at the
transmitter, the receiver and the cognitive relay, respectively.
\end{corollary}
\begin{IEEEproof}
We show that the outer bound that will be presented in Corollary~\ref{corollary:dof_mimo_outer_no} of Section~\ref{outerbound} is achievable. We consider the following three conditions:
\begin{itemize}
\itemsep 0pt
\item $2M_t +M_{\c} \leq M_r$
\item $M_t+M_{\c}\leq M_r<2M_t+M_{\c}$
\item $M_r <M_t+M_{\c}$
\end{itemize}
If $2M_t +M_{\c} \leq M_r$ or $M_t+M_{\c}\leq
M_r<2M_t+M_{\c}$, the optimal scheme is the same as that for the
delayed CSIT. This is because in Theorem~\ref{theorem:dof_mimo_cr}, the proposed
scheme for the delayed CSIT in Condition I is optimal but
does not use any delayed feedback information so that the optimal
DoF region can be obtained by this scheme. Therefore, if $2M_t
+M_{\c} \leq M_r$, the DoF region is determined by $d_{\a}\leq
M_t+M_{\c},\ d_{\b}\leq M_t+M_{\c}$, and $d_{\a}+d_{\b}\leq 2M_t+M_{\c}$. If
$M_t+M_{\c}\leq M_r<2M_t+M_{\c}$, the DoF region is determined by
$d_{\a}\leq M_t+M_{\c},\ d_{\b}\leq M_t+M_{\c}$, and $d_{\a}+d_{\b}\leq M_r$. On
the other hand, if $M_r <M_t+M_{\c}$, the DoF outer bound $d_{\a}+d_{\b}
\leq M_r$ is achievable via TDM, similar to the result for the SISO
case in Corollary~\ref{corollary:dof_siso_no}.

We note this corollary can also be obtained using the result of MIMO
BC without CSIT (i.e, without feedback) in~\cite{HJSV2012,VV2012_2}.
\end{IEEEproof}

\begin{figure}
\centering
    \subfigure[$M_r < \frac{M_t+M_{\c}}{2}$ case]
    {\includegraphics[width=0.34\columnwidth]{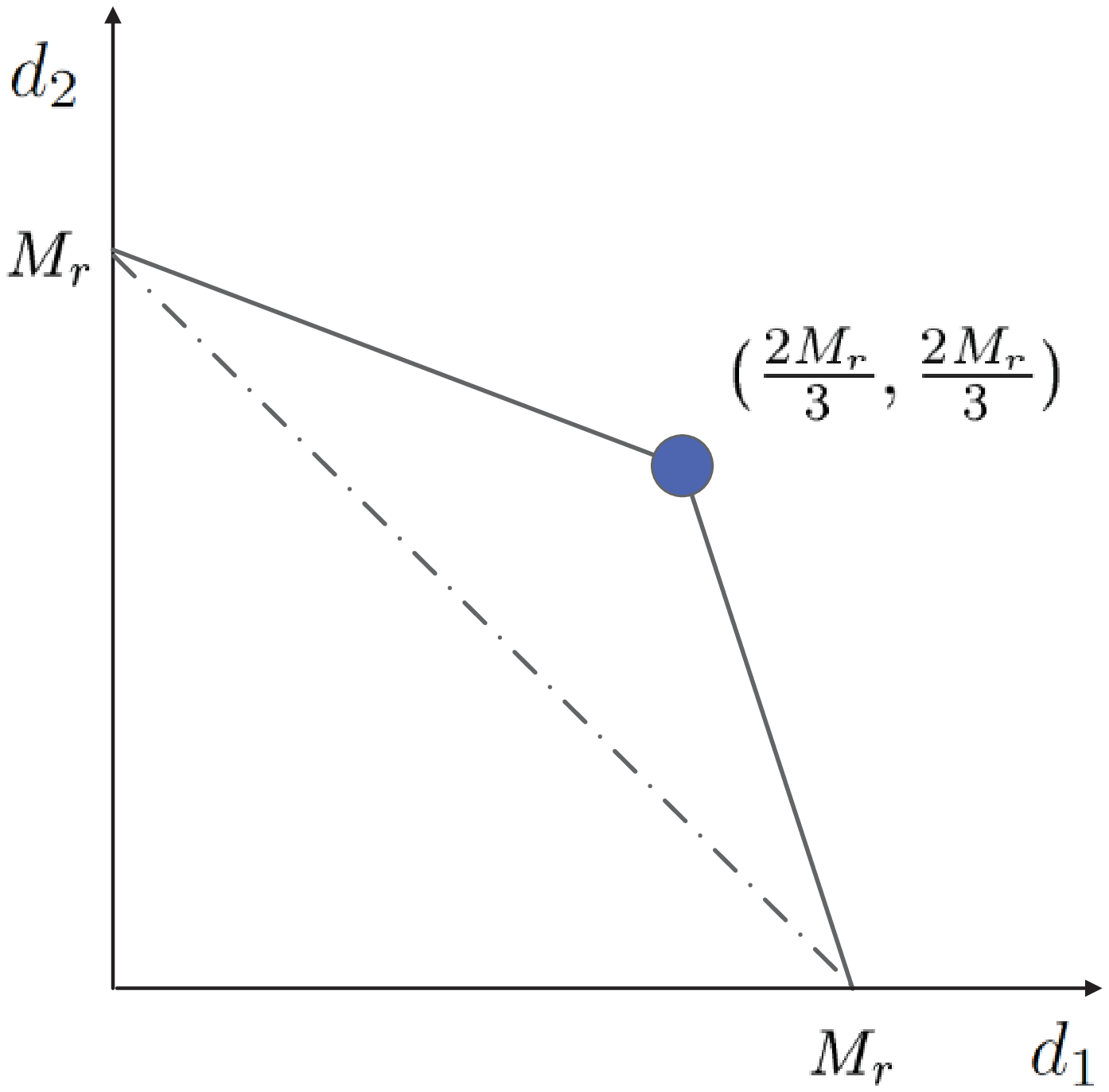}}
    \subfigure[ $\frac{M_t+M_{\c}}{2}\leq M_r <M_t+M_{\c}$ case]
    {\includegraphics[width=0.59\columnwidth]{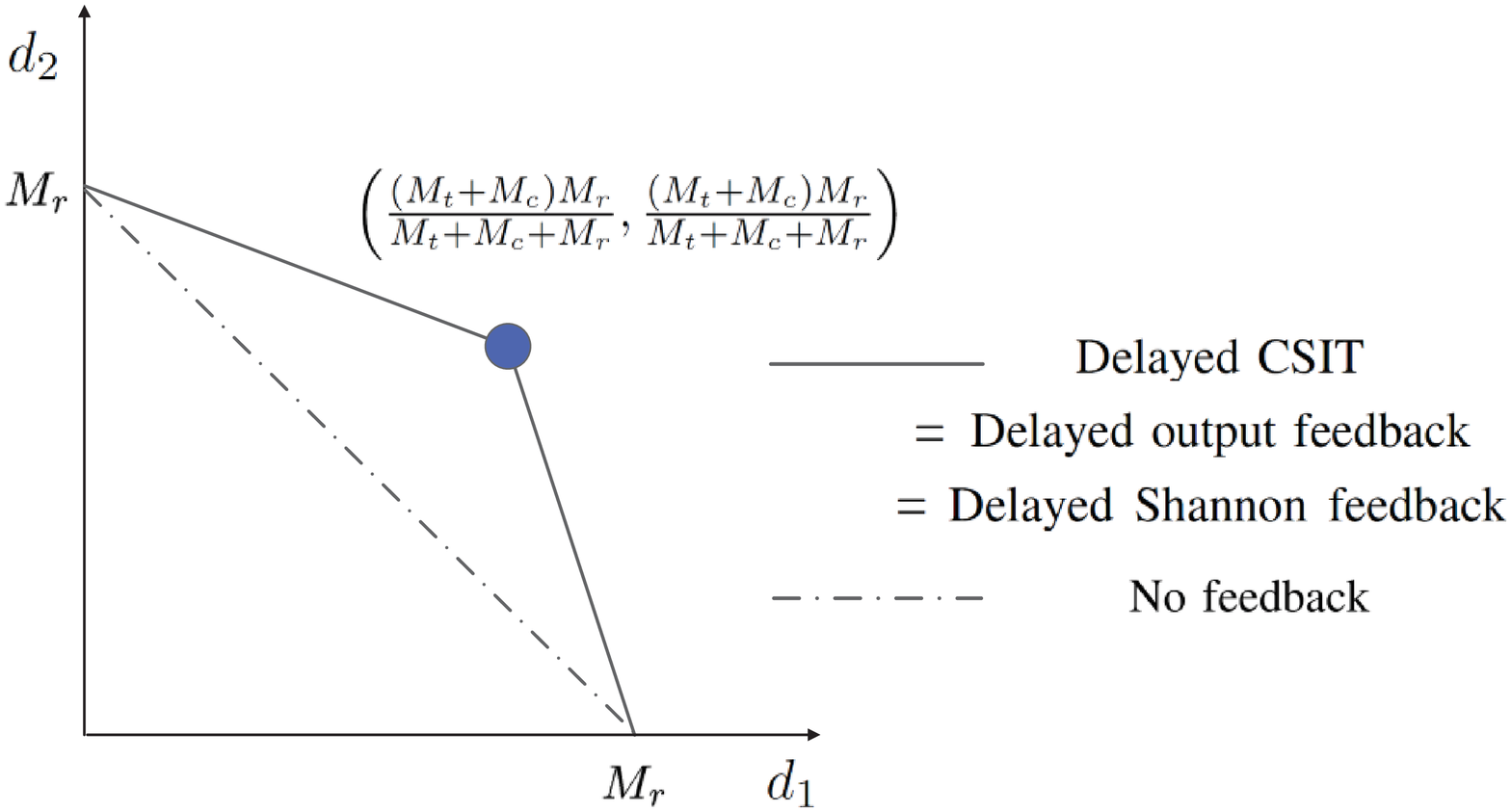}}
    \caption{The DoF region of the 
MIMO Gaussian ICCR with and without delayed feedback at both transmitters and cognitive relay.} \label{fig_dof_mimo}
\end{figure}

\begin{remark}
If $M_t+M_{\c} \leq M_r$ (i.e., Condition I), the DoF region with delayed feedback in Theorem~\ref{theorem:dof_mimo_cr} is the same as that
with no feedback. This result indicates that neither of the three types
of delayed feedback information are useful when $M_t+M_{\c} \leq M_r$.
Therefore, delayed feedback information
is useful in terms of DoF if $M_r < M_t+M_{\c}$.
Fig.~\ref{fig_dof_mimo} shows the
improvements of the DoF region by delayed feedback information at both the
transmitters and the cognitive relay, when
$M_r < \frac{M_t+M_{\c}}{2}$ and $ \frac{M_t+M_{\c}}{2} \leq
M_r<M_t+M_{\c}$, compared to the case of no feedback.
\end{remark}

\section{Achievable DoF with Delayed Feedback Unavailable at Cognitive Relay}\label{achievable_dof}
In this section, we consider the case when the transmitters
have delayed feedback information but the cognitive relay does not. Using a similar approach in Section \ref{MIMO_cr}, We derive
the achievable DoF region.

\begin{theorem} \label{theorem:dof_mimo_csit}
When the cognitive relay does not have delayed feedback information, the DoF region of the MIMO ICCR achieved by the proposed retrospective interference alignment is
\begin{align}
   \mathcal{D'}_{\mathrm{delay\setminus CR}}\subset \mathcal{\bar{D}}_{\mathrm{delay}},\quad &\mathrm{if}~M_t<M_r<M_t+M_{\c}, \nonumber\\
   \mathcal{D'}_{\mathrm{delay\setminus CR}}= \mathcal{\bar{D}}_{\mathrm{delay}},\quad & \mathrm{otherwise,} \nonumber
\end{align}

where $M_t, ~M_r$, and $M_{\c}$ are the numbers of antennas at the
transmitter, the receiver and the cognitive relay, respectively.
\end{theorem}
\begin{IEEEproof}
The DoF outer bound that will be derived in Theorem~\ref{theorem:dof_mimo_outer}
of Section~\ref{outerbound} is also valid
when delayed feedback information is not available at the cognitive
relay. We already showed that with delayed feedback, the DoF region
is achieved under Conditions I, III, and V even if the
cognitive relay does not have any feedback information. Thus, we
consider only the two cases of Conditions II and IV.

With delayed CSIT under Condition II, a DoF outer bound with
delayed CSIT for this case is given by
$\frac{d_{\a}}{M_t+M_{\c}}+\frac{d_{\b}}{M_r}\leq 1$ and
$\frac{d_{\a}}{M_r}+\frac{d_{\b}}{M_t+M_{\c}}\leq 1$. If $2M_t\geq M_r$,
we can show that the
$\Big(\frac{(M_t+M_{\c})M_t}{3M_t+M_{\c}-M_r},\frac{(M_t+M_{\c})M_t}{3M_t+M_{\c}-M_r}\Big)$
DoF pair is achievable but it does not meet the DoF outer bound.
All transmissions are scaled to satisfy the power
constraint. First, we spend $2M_t$ time slots. At each time index
$t\in \{1,...,M_t\}$, Transmitter~$\a$ sends random linear
combinations of $M_t+M_{\c}$ independent symbols, and the cognitive
relay sends distinct random linear combinations of the $M_t+M_{\c}$
independent symbols. Thus, Receiver~$\a$ has $M_r M_t$ independent
linear equations of the $(M_t+M_{\c})M_t$ desired symbols. At each
time index $t\in \{M_t+1,...,2M_t\}$, Transmitter~$\b$ and the
cognitive relay send distinct random linear combinations of
$M_t+M_{\c}$ symbols intended for Receiver~$\b$. Receiver~$\b$
obtains $M_r M_t$ independent linear combinations of the
$(M_t+M_{\c})M_t$ desired variables.
Finally, we need another $M_t+M_{\c}-M_r$ time slots
$t\in\{2M_t+1,\cdots, 3M_t+M_{\c}-M_r\}$, when Transmitter~$\a$ and Transmitter~$\b$ transmit
$X_{\a,t}=[Y_{\b[t-2M_t],1},\cdots,Y_{\b[t-2M_t],M_t}]^T$ and
$X_{\b,t}=[Y_{\a[t-2M_t],M_t+1},\cdots,Y_{\a[t-2M_t],2M_t}]^T$,
respectively, using delayed CSI, but the cognitive relay is silent.
Since the interfering terms are
comprised of the past received signal in previous slots, each
receiver can eliminate the interfering terms and obtain
$(M_t+M_{\c}-M_r)M_t$ linearly independent interference-free signals
at $t\in\{2M_t+1,\cdots, 3M_t+M_{\c}-M_r\}$. At the end of
transmission, each receiver has total $(M_t+M_{\c})M_t
(=M_rM_t+(M_t+M_{\c}-M_r)M_t)$ linearly independent equations
involving $(M_t+M_{\c})M_t$ symbols during $3M_t+M_{\c}-M_r
(=2M_t+(M_t+M_{\c}-M_r))$ time slots. Therefore, we can obtain the
$\Big(\frac{(M_t+M_{\c})M_t}{3M_t+M_{\c}-M_r},\frac{(M_t+M_{\c})M_t}{3M_t+M_{\c}-M_r}\Big)$
DoF pair, and total $\frac{2(M_t+M_{\c})M_t}{3M_t+M_{\c}-M_r}$ DoF
when $2M_t\geq M_r$.
If $2M_t< M_r$, the sum DoF $\frac{2(M_t+M_{\c})M_t}{3M_t+M_{\c}-M_r}$ achieved by the
proposed scheme is less than $M_r$, but $M_r$ is achievable via time
sharing. Thus, if you adopt time sharing instead of the proposed
scheme when $2M_t< M_r$, total $M_r$ DoF can be achievable. The other
points on the boundary of the achievable region can be obtained via
time sharing.

For Condition IV (i.e., $M_r <\frac{M_t+M_{\c}}{2}$ and $M_r > M_t$),
we can show that the $(\frac{M_t+M_r}{3},\frac{M_t+M_r}{3})$ DoF
pair is achievable, but it is below the outer bound determined by
$\frac{d_{\a}}{2M_r}+\frac{d_{\b}}{M_r}\leq 1$ and $
\frac{d_{\a}}{M_r}+\frac{d_{\b}}{2M_r}\leq 1$ if $2M_t\leq M_r$. All
transmissions are scaled to satisfy the power constraint. First, we
spend two time slots. At time index $t=1$, Transmitter~$\a$ sends $M_t$ random
linear combinations of $M_t+M_r$ symbols with $M_t$ transmit
antennas, and the cognitive relay sends $M_r$ different random
linear combinations of the $M_t+M_r$ symbols with $M_r$ antennas.
Similarly, at time index $t=2$, Transmitter~$\b$ sends $M_t$ random
linear combinations of $M_t+M_r$ symbols intended for Receiver~$\b$, and the
cognitive relay sends different $M_r$ random linear combinations of
the $M_t+M_r$ symbols. Receiver~$\b$ obtains $M_r$ linear combinations
of intended $M_t+M_r$ variables.
Second, we need one time slot indexed by $t=3$. Transmitter~$\a$ and Transmitter~$\b$
only transmit $X_{\a,3}=[Y_{\b[1],1},\cdots,Y_{\b[M_t],1}]^T$ and
$X_{\b,3}=[Y_{\a[1],2},\cdots,Y_{\a[M_t],2}]^T$, respectively, using
delayed CSI, and the cognitive relay does not transmit.
Since the interference terms at each receiver are comprised of
the received signals in the previous time slots,
each receiver can obtain $M_t$ linearly independent
interference-free signals at $t=3$. Thus, each receiver has total
$M_t+M_r$ linearly independent equations involving $M_t+M_r$ symbols so that
we can obtain the $(\frac{M_t+M_r}{3},\frac{M_t+M_r}{3})$ DoF pair and total
$\frac{2(M_t+M_r)}{3}$ DoF.
If $2M_t<M_r$, the sum DoF $\frac{2(M_t+M_r)}{3}$ achieved by the proposed
scheme is less than $M_r$, but $M_r$ can be achieved by time sharing.
Therefore, we can adopt time sharing if $2M_t<M_r$ instead of the
proposed scheme. Then, the achievable sum DoF is $M_r$ when $2M_t<M_r$.
The other points on the boundary of the achievable region are
achievable via time sharing.

Similarly, the DoF region of the 
 ICCR with delayed
output feedback can be obtained for Condition II and IV. In the
second part, the transmitters send the outputs fed back from the receivers
instead of using delayed CSI. Then, we can obtain the same DoF region
as that with delayed CSIT.
Since Shannon feedback includes CSI and output feedback and the
achievable DoF regions with delayed CSIT and delayed feedback
are identical, with delayed Shannon feedback, the same achievable DoF region can be obtained by the scheme utilizing delayed
CSIT or output feedback information.
\end{IEEEproof}

\begin{remark}
For Condition II and IV, i.e., $M_t<M_r<M_t+M_{\c}$,
the achievable DoF pairs do not meet the outer bound when the delayed
feedback information is not available at the cognitive relay. Therefore,
the proposed retrospective scheme is optimal except when
$M_t<M_r<M_t+M_{\c}$, in which case no statement about optimality can
be made at this time.
\end{remark}

\begin{remark}
Comparing Theorem~\ref{theorem:dof_mimo_csit} with Corollary~\ref{theorem:dof_mimo_no},
delayed feedback information at only transmitters is useful
in terms of DoF only if $\frac{M_t+M_{\c}}{2} \leq M_r < \min(M_t+M_{\c},2M_t)$
or $M_r < \min\left(\frac{M_t+M_{\c}}{2},2M_t\right) $.
\end{remark}

\begin{figure}
\centering
    \subfigure[$\frac{M_t+M_{\c}}{2}\leq M_t$]
    {\includegraphics[width=0.65\columnwidth]{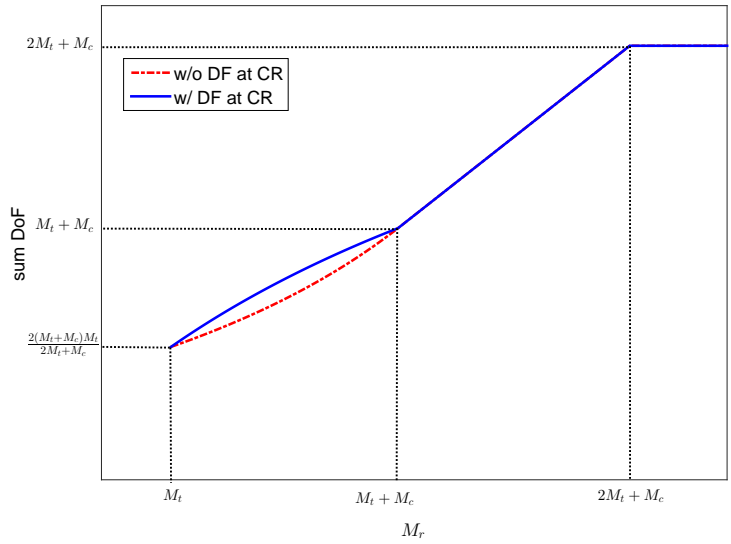}}
    \subfigure[$M_t<\frac{M_t+M_{\c}}{2}<2M_t$]
    {\includegraphics[width=0.65\columnwidth]{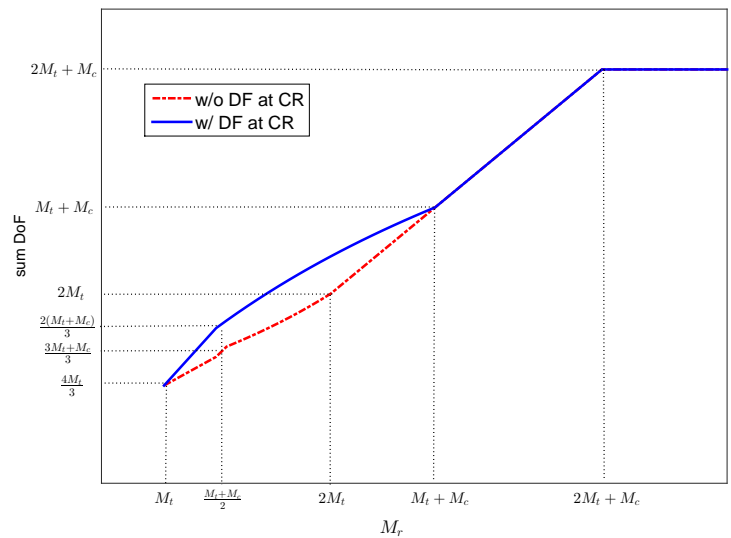}}
    \caption{The achievable sum DoF when $M_r >M_t$ for fixed $M_t$ and $M_{\c}$.} \label{fig_dof_com}
\end{figure}

\begin{remark}
Fig.~\ref{fig_dof_com} shows the achievable sum DoFs for the two cases
with/without delayed feedback at the cognitive relay when $M_r >M_t$
for fixed $M_t$ and $M_{\c}$. Fig.~\ref{fig_dof_com}(a) corresponds to Condition I and II when $\frac{M_t+M_{\c}}{2}\leq M_t$,
and Fig.~\ref{fig_dof_com}(b) corresponds to Condition I, II and IV when $M_t<\frac{M_t+M_{\c}}{2}<2M_t$.
\end{remark}

\section{DoF Outer Bounds}
\label{outerbound}

\subsection{Delayed feedback}
\label{delayed_outer}
The following outer bound holds for all three types of feedback information discussed in this paper.

\begin{theorem} \label{theorem:dof_mimo_outer}
The DoF region  with
delayed feedback is contained in the following region:
\begin{align}
   \mathcal{\bar{D}}_\mathrm{delayed}= \Bigg\{(d_{\a},d_{\b})\in\mathbb{R}_{+}^2\ :\ &\ \ \ \ \ \ \  d_{\a}\leq \min(M_r,M_t+M_{\c}),\ \  d_{\b}\leq \min(M_r,M_t+M_{\c}), \nonumber\\
   & \frac{d_{\a}}{\min(M_r,M_t+M_{\c})}+\frac{d_{\b}}{\min(2M_r,M_t+M_{\c})}\leq \frac{\min(M_r,2M_t+M_{\c})}{\min(M_r,M_t+M_{\c})},\nonumber\\
   & \frac{d_{\a}}{\min(2M_r,M_t+M_{\c})}+\frac{d_{\b}}{\min(M_r,M_t+M_{\c})}\leq \frac{\min(M_r,2M_t+M_{\c})}{\min(M_r,M_t+M_{\c})} \Bigg\} \nonumber
\end{align}
where $M_t, ~M_r$, and $M_{\c}$ are the numbers of antennas at the
transmitter, the receiver and the cognitive relay, respectively.
\end{theorem}

\begin{lemma} \label{lemma_mimo1}
For a given $t\in\{1,2,\cdots,n\}$,
    \begin{align}
        \frac{1}{c_1}h(Y_{\a[1:c_1],t}|Y_{\a[1:M_r]}^{t-1},W_{\a},\mathcal{H}^n)\geq \frac{1}{c_2}h(Y_{\a[1:c_1],t},Y_{\b[1:c_2-c_1],t}|Y_{\a[1:M_r]}^{t-1},Y_{\b[1:M_r]}^{t-1},W_{\a},\mathcal{H}^n). \nonumber
    \end{align}
where $c_1 \triangleq \min(M_r,M_t+M_{\c})$ and $c_2 \triangleq \min(2M_r,M_t+M_{\c})$.
\end{lemma}
\begin{IEEEproof}
The key idea of this proof is the statistical equivalence of channel outputs~\cite{VV2012, TMPS2013}. The detailed proof is in Appendix~\ref{proof_1}.
\end{IEEEproof}

\begin{lemma} \label{lemma_mimo2}
For the 
 ICCR with delayed feedback information, we
have
    \begin{align}
        \frac{1}{\min(M_r,M_t+M_{\c})}h(Y_{\a[1:M_r]}^n|W_{\a},\mathcal{H}^n)\geq \frac{1}{\min(2M_r,M_t+M_{\c})}h(Y_{\a[1:M_r]}^n,Y_{\b[1:M_r]}^n|W_{\a},\mathcal{H}^n)+n\cdot o(\mathrm{log}_2P). \nonumber
    \end{align}
\end{lemma}
\begin{IEEEproof}
We use Lemma~\ref{lemma_mimo1} to prove Lemma~\ref{lemma_mimo2}. The detailed proof is in Appendix~\ref{proof_2}.
\end{IEEEproof}\vspace{0.1in}
Using Lemma~\ref{lemma_mimo1} and Lemma~\ref{lemma_mimo2}, we now prove
Theorem~\ref{theorem:dof_mimo_outer}.

\begin{IEEEproof}
The outer bounds $d_{\a}\leq \min(M_r,M_t+M_{\c})$ and $d_{\b}\leq
\min(M_r,M_t+M_{\c})$ can be readily obtained from the numbers of
antennas. The other bounds are obtained using the fact that the
conditional distributions of $Y_{\a[\ell_1],t}$ and $Y_{\b[\ell_2],t}$
for all $\ell_1,~\ell_2\in\{1,\cdots,M_r\}$ are identical when the
two variables are conditioned on the collection of channel gains
over all time, past channel outputs, and some present channel
outputs.

For block length $n$, using Fano's inequality we can bound the rate
$R_{\a}$ as
\begin{align}
    nR_{\a}&\leq I(W_{\a};Y_{\a[1:M_r]}^n|\mathcal{H}^n)+n\varepsilon_{\a,n} \nonumber \\
    &=h(Y_{\a[1:M_r]}^n|\mathcal{H}^n)-h(Y_{\a[1:M_r]}^n|W_{\a},\mathcal{H}^n)+n\varepsilon_{\a,n}   \nonumber \\
    &\leq n\min(M_r,2M_t+M_{\c})\mathrm{log}_2P-h(Y_{\a[1:M_r]}^n|W_{\a},\mathcal{H}^n)+n\cdot o(\mathrm{log}_2P)+n\varepsilon_{\a,n} \label{mimoouter1}
\end{align}
where $\varepsilon _{\a,n}\!\rightarrow 0$ as $n\!\rightarrow 0$.

For the rate $R_{\b}$, we obtain an outer bound using Fano's inequality
as
\begin{align}
    nR_{\b}&\leq I(W_{\b};Y_{\b[1:M_r]}^n|\mathcal{H}^n)+n\varepsilon_{\b,n} \nonumber \\
    &\leq I(W_{\b};Y_{\b[1:M_r]}^n,Y_{\a[1:M_r]}^n|W_{\a},\mathcal{H}^n)+n\varepsilon_{\b,n}   \nonumber \\
    &=h(Y_{\a[1:M_r]}^n,Y_{\b[1:M_r]}^n|W_{\a},\mathcal{H}^n)\!-\!h(Y_{\a[1:M_r]}^n,Y_{\b[1:M_r]}^n|W_{\a},W_{\b},\mathcal{H}^n)+n\varepsilon_{\b,n}  \nonumber \\
    &\leq h(Y_{\a[1:M_r]}^n,Y_{\b[1:M_r]}^n|W_{\a},\mathcal{H}^n)+n\varepsilon_{\b,n}   \label{mimoouter2}
\end{align}
where $\varepsilon _{\b,n}\!\rightarrow 0$ as $n\!\rightarrow 0$.

By Lemma~\ref{lemma_mimo2}, we can combine (\ref{mimoouter1}) and
(\ref{mimoouter2}) as
\begin{align}
    \frac{nR_{\a}}{\min(M_r,M_t+M_{\c})}+\frac{nR_{\b}}{\min(2M_r,M_t+M_{\c})}\leq \frac{n\min(M_r,2M_t+M_{\c})}{\min(M_r,M_t+M_{\c})}\log_2 P +n\cdot o(\mathrm{log}_2P)+n\varepsilon_{n}  \nonumber
\end{align}
where $\varepsilon_{n}=\varepsilon_{\a,n}+\varepsilon_{\b,n}\rightarrow 0$ as
$n\!\rightarrow 0$. Hence, we obtain DoF outerbound as
\begin{align}
    \frac{d_{\a}}{\min(M_r,M_t+M_{\c})}+\frac{d_{\b}}{\min(2M_r,M_t+M_{\c})}\leq \frac{\min(M_r,2M_t+M_{\c})}{\min(M_r,M_t+M_{\c})}. \nonumber
\end{align}
Similarly, we can obtain
\begin{align}
    \frac{d_{\a}}{\min(2M_r,M_t+M_{\c})}+\frac{d_{\b}}{\min(M_r,M_t+M_{\c})}\leq \frac{\min(M_r,2M_t+M_{\c})}{\min(M_r,M_t+M_{\c})} \nonumber
\end{align}
by switching the receiver order.
\end{IEEEproof}

\subsection{No feedback} \label{outer_no}

A DoF outer bound in the absence of CSIT can be obtained in a straight forward manner using the results
from~\cite{HJSV2012,VV2012_2}.
\begin{corollary}\label{corollary:dof_mimo_outer_no}
The outer bound of the DoF region with no feedback
$\mathcal{\bar{D}}_{no}$ is
    \begin{align}
        \mathcal{\bar{D}}_{no}= \Big\{(d_{\a},d_{\b})\in\mathbb{R}_{+}^2\; : & \; d_{\a}\leq \min(M_t+M_{\c},M_r), 
         \; d_{\b}\leq \min(M_t+M_{\c},M_r),\nonumber\\
        &\; d_{\a}+d_{\b}\leq \min(2M_t+M_{\c},M_r) \Big\}. \label{dof_mimo_no_out}
    \end{align}
where $M_t,~M_r$, and $M_{\c}$ are the numbers of antennas at the
transmitter, the receiver, and the cognitive relay, respectively.
\end{corollary}
\begin{IEEEproof}
Consider a transmitter-cooperative outer bound. Because the
transmitter cooperation results in the MIMO broadcast channel with $2M_t+M_{\c}$
transmit antennas and two receivers with $M_r$-antenna each. The DoF
outer bound follows directly from the results
of~\cite{HJSV2012,VV2012_2}.
\end{IEEEproof}

\section{Discussions}\label{discussion}
\subsection{Comparisons with Broadcast and Interference Channel with delayed CSIT}

If cooperation among transmitters and cognitive relay is allowed, the
 ICCR becomes equivalent to the 
 broadcast
channel where the transmitter has $2M_t+M_{\c}$ antennas and each
receiver has $M_r$ antennas. Therefore, when CSIT is delayed, the DoF
region of the 
 broadcast channel with antenna configuration
$(2M_t+M_{\c}, M_r, M_r)$ is a superset of the DoF of the ICCR under
$(M_t, M_t, M_{\c}, M_r, M_r)$. Table~\ref{table2} shows a comparison
of the sum-DoF under delayed CSIT between a 
 broadcast
channel~\cite{VV2011} and ICCR where delayed CSIT is available at all
nodes.
If
$2M_t+M_{\c}\leq M_r$ or $M_r<\frac{M_t+M_{\c}}{2}$, the sum DoF is
the same. For the other scenarios (i.e., $M_t+\frac{M_{\c}}{2}< M_r
<2M_t+M_{\c}$), the sum DoF of the MIMO ICCR with delayed
CSIT is less than that of the MIMO broadcast channel.

\begin{table}
\caption{Sum DoFs for MIMO broadcast channel, ICCR and interference channel with delayed CSIT}
\label{table2}
\begin{center}
\begin{tabular}{|c |c |c |c |}
\hline & broadcast channel~\cite{VV2011} & ICCR & interference channel (for even $M_{\c}$)~\cite{VV2012} \\ \hline\hline $2M_t+M_{\c}\leq M_r$ &$2M_t+M_{\c}$
&$2M_t+M_{\c}$ &$2M_t+M_{\c}$ \\ \hline $M_t+M_{\c}\leq
M_r<2M_t+M_{\c}$ &$\frac{2(2M_t+M_{\c})M_r}{2M_t+M_{\c}+M_r}$ &$M_r$ &$M_r$ \\
\hline $M_t+\frac{M_{\c}}{2}\leq M_r<M_t+M_{\c}$
&$\frac{2(2M_t+M_{\c})M_r}{2M_t+M_{\c}+M_r}$ &$\frac{2(M_t+M_{\c})M_r}{M_t+M_{\c}+M_r}$ &$M_r$ \\
\hline $\frac{M_t+M_{\c}}{2}\leq M_r<M_t+\frac{M_{\c}}{2}$
&$\frac{4M_r}{3}$& $\frac{2(M_t+M_{\c})M_r}{M_t+M_{\c}+M_r}$
&$\frac{2(M_t+\frac{M_{\c}}{2})M_r}{M_t+\frac{M_{\c}}{2}+M_r}$\\
\hline $M_r<\frac{M_t+M_{\c}}{2}$ & $\frac{4M_r}{3}$ &
$\frac{4M_r}{3}$ &$\frac{4M_r}{3}$\\ \hline
\end{tabular}
\end{center}
\end{table}

Table~\ref{table2} reproduces from~\cite{VV2012} the sum-DoF of the
MIMO interference channel, with $M_t+\frac{M_{\c}}{2}$ transmit and
$M_r$ receive antennas at respective nodes. If the two transmitters
partially cooperate, the channel becomes equivalent to the ICCR with
antenna configuration of $(M_t, M_t, M_{\c}, M_r, M_r)$. Therefore,
the DoF region of the interference channel with delayed CSIT is
included by that of the ICCR with delayed CSIT. If $M_t+M_{\c}\leq
M_r$ or $M_r<\frac{M_t+M_{\c}}{2}$, the sum DoF is the same for both
channels. If $\frac{M_t+M_{\c}}{2}\leq M_r \leq M_t+M_{\c}$, however,
the sum DoF of the ICCR is greater than that of the 
interference channel because the cognitive relay effectively produces
partial cooperation between transmitters.

\subsection{Extension to  
 Cognitive Interference Channel}\label{application}

\begin{figure}
\centering
    \includegraphics[width=0.8\columnwidth]{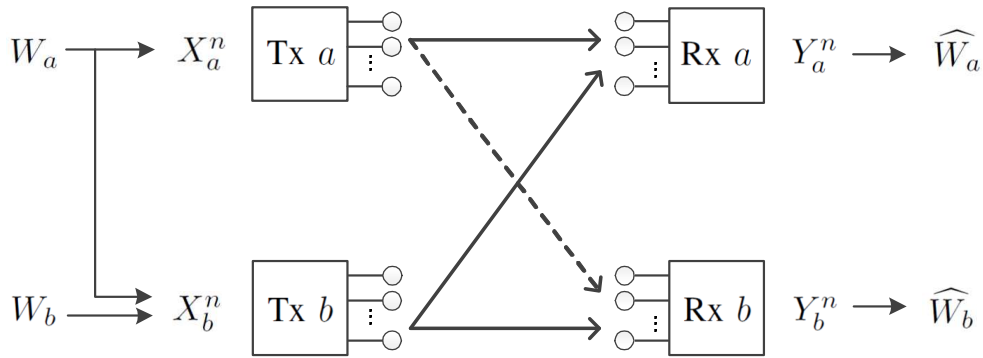}
    \caption{A 
MIMO cognitive interference channel.} \label{cognitiveIC}
\end{figure}

Here we consider another extension to the 
 cognitive interference channel
consisting of one non-cognitive transmitter, one cognitive
transmitter, and their intended receivers. The cognitive transmitter
has both messages intended for the two receivers as shown in Fig.~\ref{cognitiveIC}. The 
 cognitive interference channel with perfect CSIT and
CSIR has been studied in~\cite{DMT2006,MYK2007,MGKS2008,RG2013,HJ2009,RG2014}. The inner and outer
bounds of capacity region of the 
 SISO cognitive interference channel with
perfect CSIT were given in~\cite{DMT2006,MYK2007,MGKS2008}. For MIMO
cognitive interference channel,~\cite{RG2013,RG2014} calculated the capacity
region within a constant gap. In~\cite{HJ2009}, the
DoF region of the 
 cognitive interference channel with perfect CSIT was
derived. However, the DoF region of the 
 cognitive interference channel with
delayed feedback has been unknown. The DoF of the
 ICCR from the previous section can be used for a lower
and an upper bound of the 
 cognitive interference channel when feedback
is delayed.
\begin{corollary} \label{corollary:cognitiveIC_lower}
The DoF region of the 
 cognitive interference channel with antenna
configuration  $(M_t, M_t+M_{\c}, M_r, M_r)$ is lower bounded by
that of the 
 ICCR with antenna configuration
$(M_t,M_t ,M_{\c},M_r,M_r)$.
\end{corollary}
\begin{IEEEproof}
If cooperation between the cognitive relay and one transmitter is
allowed in the 
 ICCR, the channel becomes equivalent
to the 
 cognitive interference channel where the non-cognitive
transmitter has $M_t$ antennas, the cognitive transmitter has $M_t +
M_{\c}$ antennas, and each receiver has $M_r$ antennas. Therefore,
the DoF region of the 
 cognitive interference channel with antenna
configuration  $(M_t, M_t+M_{\c}, M_r, M_r)$ is lower bounded by
that of the 
 ICCR with antenna configuration
$(M_t,M_t ,M_{\c},M_r,M_r)$.
\end{IEEEproof}
\begin{corollary} \label{corollary:cognitiveIC_upper}
The DoF region of the 
 cognitive interference channel with antenna configuration
$(M_t, M_{\c}, $ $M_r, M_r)$ is upper bounded by the DoF region of
the 
 ICCR with antenna configuration $(M_t,M_t
,M_{\c},M_r,M_r)$.
\end{corollary}
\begin{IEEEproof}
If only one transmitter exists in the 
 ICCR, the
antenna configuration for this scenario is $(M_t,0,M_{\c},M_r,M_r)$
and it corresponds to the 
 cognitive interference channel where the
non-cognitive transmitter has $M_t$ antennas and the cognitive
transmitter has $M_{\c}$ antennas while each receiver has $M_r$
antennas. Hence, the upper bound of the DoF region of the cognitive
interference channel with antenna configuration of $(M_t, M_{\c}, $ $M_r, M_r)$ is
that of the 
 ICCR with antenna configuration of
$(M_t,M_t,M_{\c},M_r,M_r)$.
\end{IEEEproof}

\begin{figure}
\centering
    \includegraphics[width=0.6\columnwidth]{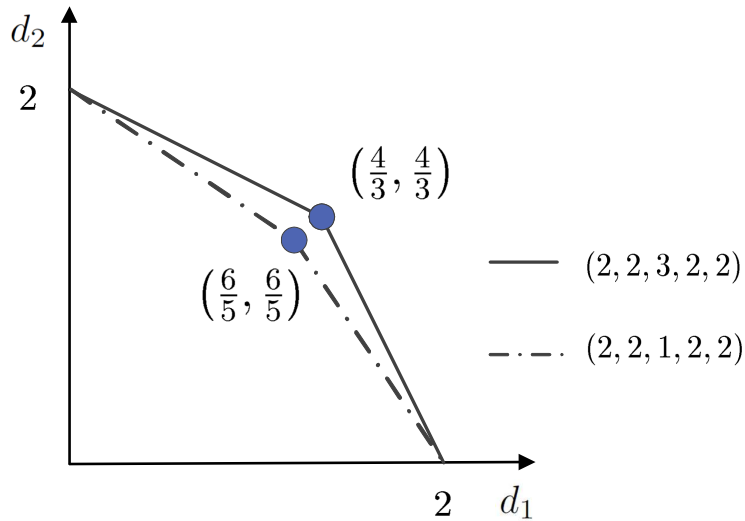}
    \caption{The DoF regions of the 
ICCR with delayed feedback for $(2,2,1,2,2)$ and $(2,2,3,2,2)$.} \label{dof_cognitive}
\end{figure}

\begin{example}[
 Cognitive interference channel with antenna configuration $(2, 3, 2, 2)$]
In this example the non-cognitive transmitter has
two antennas, the cognitive transmitter has three antennas, and
receivers have two antennas each.

The DoF region of the $(2,2,1,2,2)$ ICCR can serve as a lower bound, and the DoF of $(2,2,3,2,2)$ ICCR serves as upper bound. The lower bound derived in Section~\ref{MIMO_cr} is
\begin{align}
   \mathcal{D}_\mathrm{CSI}= \bigg\{(d_{\a},d_{\b})\in\mathbb{R}_{+}^2\ : \ \frac{d_{\a}}{2}+\frac{d_{\b}}{3}\leq 1,\ \ \frac{d_{\a}}{3}+\frac{d_{\b}}{2}\leq 1 \bigg\}, \label{dof_(2,2,1,2,2)}
\end{align} where the maximum sum DoF is $\frac{12}{5}$.
The upper bound is obtained from the results of Section~\ref{MIMO_cr} as
\begin{align}
   \mathcal{D}_\mathrm{CSI}= \bigg\{(d_{\a},d_{\b})\in\mathbb{R}_{+}^2\ : \ \frac{d_{\a}}{2}+\frac{d_{\b}}{4}\leq 1,\ \ \frac{d_{\a}}{4}+\frac{d_{\b}}{2}\leq 1 \bigg\}, \label{dof_(2,2,3,2,2)}
\end{align} where the maximum sum DoF is
$\frac{8}{3}$. These lower and upper bounds are shown in Fig.~\ref{dof_cognitive}.
\end{example}

\section{Conclusion}\label{conclusion}

This paper studies the DoF region of the two-user
 Gaussian fading interference channel with cognitive relay
(ICCR) with delayed feedback. Three different types of delayed
feedback are considered: delayed CSIT, delayed output feedback, and delayed
Shannon feedback. For the
 SISO ICCR, the proposed retrospective interference alignment
scheme using delayed feedback information achieves the DoF region.
The sum DoF of the 
 SISO ICCR is
$4/3$ with delayed feedback information, compared to the DoF of 1
for SISO interference channel in the absence of relay, regardless of CSIT.
Without feedback, the cognitive relay is not useful in the sense of DoF.

In the MIMO case, the optimal DoF has been characterized under all
antenna configurations if delayed feedback is provided to
both the transmitters and cognitive relay. DoF benefits can be obtained
over and above the open-loop system when $\frac{M_t+M_{\c}}{2} \leq
M_r < M_t+M_{\c} $ or $M_r < \frac{M_t+M_{\c}}{2}$, and the sum DoF is
$\frac{2(M_t+M_{\c})M_r}{M_t+M_{\c}+M_r}$ when $\frac{M_t+M_{\c}}{2}
\leq M_r < M_t+M_{\c} $ and $\frac{4M_r}{3}$ when $M_r <
\frac{M_t+M_{\c}}{2}$. On the other hand, if $2M_t +M_{\c} \leq M_r$
or $M_t+M_{\c} \leq M_r < 2M_t+M_{\c}$, then delayed feedback does not
help in the sense of DoF. In this scenario, the sum-DoF (for both
open-loop and closed-loop) is $2M_t+M_{\c}$ when $2M_t +M_{\c}\leq
M_r$ and $M_r$ when $M_t+M_{\c} \leq M_r < 2M_t+M_{\c}$.

If delayed feedback is unavailable at the cognitive relay,
the proposed retrospective interference alignment
scheme achieves the optimal DoF except when $M_t<M_r<M_t+M_{\c}$,
where existing upper and lower bounds do not meet. Delayed feedback
is shown to extend the DoF over and above the open-loop system when
$(M_t+M_{\c})/2 \leq M_r < \min\big(M_t+M_{\c},2M_t\big)$ and $M_r <
\min\big((M_t+M_{\c})/2,2M_t\big)$.

In addition, in this paper upper and lower bounds are derived for the DoF
region of the two-user MIMO cognitive
interference channel under delayed feedback.

\appendices
\def\thesection{\Alph{section}}%
\def\thesectiondis{\Alph{section}}%

\section{Proof of Lemma 1} \label{proof_1}
\setcounter{equation}{0}
\renewcommand{\theequation}{A.\arabic{equation}}

\begin{align}
     \frac{1}{c_1}h(Y_{\a[1:c_1],t}|Y_{\a[1:M_r]}^{t-1},W_{\a},\mathcal{H}^n)
     &=\frac{1}{c_1}\sum_{\ell=1}^{c_1}h(Y_{\a[\ell],t}|Y_{\a[1:\ell-1],t},Y_{\a[1:M_r]}^{t-1},W_{\a},\mathcal{H}^n)\nonumber\\
     &\stackrel{(\mathrm{a})}{\geq} \frac{1}{c_1}\sum_{\ell=1}^{c_1}h(Y_{\a[\ell],t}|Y_{\a[1:c_1-1],t},Y_{\a[1:M_r]}^{t-1},W_{\a},\mathcal{H}^n)\label{a1}
\end{align}
where (a) holds because conditioning reduces entropy.
Since $X_{\a,t}=f_{\a,t}(W_{\a},\mathcal{H}^{t-1})$ if only
delayed CSIT is available, we can write
\begin{align}
        h(Y_{\a[\ell_1],t}|Y_{\a[1:c_1-1],t},Y_{\a[1:M_r]}^{t-1},W_{\a},\mathcal{H}^n) &=h(Y_{\a[\ell_1],t}|X_{\a,t}, Y_{\a[1:c_1-1],t},Y_{\a[1:M_r]}^{t-1},W_{\a},\mathcal{H}^n) \nonumber\\
        &=h(\widetilde{Y}_{\a[\ell_1],t}|Y_{\a[1:c_1-1],t},Y_{\a[1:M_r]}^{t-1},W_{\a},\mathcal{H}^n)\label{entropy_y_1i}\\
        h(Y_{\b[\ell_2],t}|Y_{\a[1:c_1-1],t},Y_{\a[1:M_r]}^{t-1},W_{\a},\mathcal{H}^n) &=h(Y_{\b[\ell_2],t}|X_{\a,t}, Y_{\a[1:c_1-1],t},Y_{\a[1:M_r]}^{t-1},W_{\a},\mathcal{H}^n) \nonumber\\
        &=h(\widetilde{Y}_{\b[\ell_2],t}|Y_{\a[1:c_1-1],t},Y_{\a[1:M_r]}^{t-1},W_{\a},\mathcal{H}^n)\label{entropy_y_2i}
    \end{align}
where $\widetilde{Y}_{i[\ell],t}$ is defined as the output signal if
we assume $X_{\a,t}=\mathbf{0}_{M_t\times 1}$, i.e., in the absence of
$X_{\a,t}$.  We note that $\widetilde{Y}_{\a[\ell_1],t}$ and
$\widetilde{Y}_{\b[\ell_2],t}$ have identical distributions since the
respective channel gains and receiver noises have identical
distributions. Thus, for all $\ell_1$ and $\ell_2$,
\begin{align}
    h(\widetilde{Y}_{\a[\ell_1],t}|Y_{\a[1:c_1-1],t},Y_{\a[1:M_r]}^{t-1},W_{\a},\mathcal{H}^n)=h(\widetilde{Y}_{2[\ell_2],t}|Y_{\a[1:c_1-1],t},Y_{\a[1:M_r]}^{t-1},W_{\a},\mathcal{H}^n) \label{equal1}
\end{align}
therefore,
\begin{align}
     h(Y_{\a[\ell_1],t}|Y_{\a[1:c_1-1],t},Y_{\a[1:M_r]}^{t-1},W_{\a},\mathcal{H}^n)= h(Y_{\b[\ell_2],t}|Y_{\a[1:c_1-1],t},Y_{\a[1:M_r]}^{t-1},W_{\a},\mathcal{H}^n).\label{equal2}
\end{align}  This represents the
statistical equivalence of channel outputs~\cite{VV2012, TMPS2013}.
The result (\ref{equal2}) is also applicable even if
$X_{\a,t}=f_{\a,t}(W_{\a},\mathcal{H}^{t-1})$ is replaced by
$X_{\a,t}=f_{\a,t}(W_{\a},{Y}_{\a}^{t-1})$ or
$X_{\a,t}=f_{\a,t}(W_{\a},{Y}_{\a}^{t-1},\mathcal{H}^{t-1})$ for the
delayed feedback or the Shannon feedback information, since
$X_{\a,t}$ when either delayed feedback or Shannon feedback
information is available can be constructed from the given
conditions for the delayed CSIT case. Thus, we can rewrite
(\ref{a1}) as
\begin{align}
    \frac{1}{c_1}h(Y_{\a[1:c_1],t}|Y_{\a[1:M_r]}^{t-1},W_{\a},\mathcal{H}^n)&\geq \frac{1}{c_1}\sum_{\ell=1}^{c_1}h(Y_{\a[\ell],t}|Y_{\a[1:c_1-1],t},Y_{\a[1:M_r]}^{t-1},W_{\a},\mathcal{H}^n) \nonumber\\
     &\stackrel{(\mathrm{b})}{=} h(Y_{\b[1],t}|Y_{\a[1:c_1-1],t},Y_{\a[1:M_r]}^{t-1},W_{\a},\mathcal{H}^n) \nonumber\\
     &\stackrel{(\mathrm{c})}{\geq} h(Y_{\b[1],t}|Y_{\a[1:c_1],t},Y_{\a[1:M_r]}^{t-1},W_{\a},\mathcal{H}^n)\nonumber\\
     &\stackrel{(\mathrm{d})}{=}\frac{1}{c_2-c_1}\sum_{\ell=1}^{c_2-c_1}h(Y_{\b[\ell],t}|Y_{\a[1:c_1],t},Y_{\a[1:M_r]}^{t-1},W_{\a},\mathcal{H}^n)\nonumber\\
     &\stackrel{(\mathrm{e})}{\geq} \frac{1}{c_2-c_1}\sum_{\ell=1}^{c_2-c_1}h(Y_{\b[\ell],t}|Y_{\b[1:\ell-1],t},Y_{\a[1:c_1],t},Y_{\a[1:M_r]}^{t-1},W_{\a},\mathcal{H}^n)\nonumber\\
     &=\frac{1}{c_2-c_1}h(Y_{\b[1:c_2-c_1],t}|Y_{\a[1:c_1],t},Y_{\a[1:M_r]}^{t-1},W_{\a},\mathcal{H}^n)\label{a2}
\end{align}
where (b) and (d) follow from the statistical equivalence of
channel outputs, and (c) and (e) follow from the fact that conditioning reduces
entropy. Thus, we have
\begin{align}
    c_2h(Y_{\a[1:c_1],t}|Y_{\a[1:M_r]}^{t-1},W_{\a},\mathcal{H}^n)&\geq c_1h(Y_{\a[1:c_1],t},Y_{\b[1:c_2-c_1],t}|Y_{\a[1:M_r]}^{t-1},W_{\a},\mathcal{H}^n)\nonumber\\
    &\stackrel{(\mathrm{f})}{\geq}  c_1 h(Y_{\a[1:c_1],t},Y_{\b[1:c_2-c_1],t}|Y_{\a[1:M_r]}^{t-1},Y_{\b[1:M_r]}^{t-1},W_{\a},\mathcal{H}^n)\nonumber
\end{align}
where (f) follows from the fact that conditioning reduces entropy.

\section{Proof of Lemma 2} \label{proof_2}
\setcounter{equation}{0}
\renewcommand{\theequation}{B.\arabic{equation}}

We use Lemma~\ref{lemma_mimo1}, as follows:
\begin{align}
    \frac{1}{c_1}h(Y_{\a[1:M_r]}^n|W_{\a},\mathcal{H}^n)&\geq \frac{1}{c_1}h(Y_{\a[1:c_1]}^n|W_{\a},\mathcal{H}^n)\nonumber\\
    &=\frac{1}{c_1} \sum_{t=1}^{n}h(Y_{\a[1:c_1],t}|Y_{\a[1:c_1]}^{t-1},W_{\a},\mathcal{H}^n)\nonumber\\
    &\stackrel{(\mathrm{g})}{\geq}\frac{1}{c_1} \sum_{t=1}^{n}h(Y_{\a[1:c_1],t}|Y_{\a[1:M_r]}^{t-1},W_{\a},\mathcal{H}^n)\nonumber\\
    &\stackrel{(\mathrm{h})}{\geq} \frac{1}{c_2} \sum_{t=1}^{n}h(Y_{\a[1:c_1],t},Y_{\b[1:c_2-c_1],t}|Y_{\a[1:M_r]}^{t-1},Y_{\b[1:M_r]}^{t-1},W_{\a},\mathcal{H}^n)\nonumber\\
    &= \frac{1}{c_2}\sum_{t=1}^{n}\!\Big[h(Y_{\a[1:M_r],t},Y_{\b[1:M_r],t}|Y_{\a[1:M_r]}^{t-1},Y_{\b[1:M_r]}^{t-1},W_{\a},\mathcal{H}^n)\nonumber\\
    &\ \ \ \ -h(Y_{\a[c_1+1:M_r],t},Y_{\b[c_2-c_1+1:M_r],t}|Y_{\a[1:c_1],t},Y_{\b[1:c_2-c_1],t},Y_{\a[1:M_r]}^{t-1},Y_{\b[1:M_r]}^{t-1},W_{\a},\mathcal{H}^n)\Big]\nonumber\\
    &\stackrel{(\mathrm{i})}{\geq}  \frac{1}{c_2}h(Y_{\a[1:M_r]}^n,Y_{\b[1:M_r]}^n|W_{\a},\mathcal{H}^n)+n\cdot o(\mathrm{log}_2P)
\end{align}
where $c_1=\min(M_r,M_t+M_{\c})$, $c_2=\min(2M_r,M_t+M_{\c})$, (g) follows from the
fact that conditioning reduces entropy, and (h) follows from Lemma~\ref{lemma_mimo1}, (i) follows from the fact that for all $t\in\{1,\cdots,n\}$,
\begin{align}
    h&(Y_{\a[c_1+1:M_r],t},Y_{\b[c_2-c_1+1:M_r],t}|Y_{\a[1:c_1],t},Y_{\b[1:c_2-c_1],t},Y_{\a[1:M_r]}^{t-1},Y_{\b[1:M_r]}^{t-1},W_{\a},\mathcal{H}^n)\nonumber\\
    &=h(Y_{\a[c_1+1:M_r],t},Y_{\b[c_2-c_1+1:M_r],t}|Y_{\a[1:c_1],t},Y_{\b[1:c_2-c_1],t},Y_{\a[1:M_r]}^{t-1},Y_{\b[1:M_r]}^{t-1},X_{\a}^n,W_{\a},\mathcal{H}^n)\nonumber\\
    &\leq  o(\mathrm{log}_2P)\label{o_log}
\end{align}
because $Y_{\a[c_1+1:M_r],t}$ and $Y_{\b[c_2-c_1+1:M_r],t}$ do not
affect the DoF when  the channel inputs $X_{\a}^n$ and the channel
outputs $Y_{\a[1:c_1],t}$ and $Y_{\b[1:c_2-c_1]}$ are given.
Therefore, we have
\begin{align}
    \frac{1}{c_1}h(Y_{\a[1:M_r]}^n|W_{\a},\mathcal{H}^n)\geq \frac{1}{c_2}h(Y_{\a[1:M_r]}^n,Y_{\b[1:M_r]}^n|W_{\a},\mathcal{H}^n)+n\cdot o(\mathrm{log}_2P). \nonumber
\end{align}

\newpage

\end{document}